\documentstyle[psfig,mncite,epsf,subfigure,longtable,amstex,amssymb]{mn}

\begin{document}
\title[Constraints on Jupiters from observations of Galactic Bulge microlensing events during 2000.]{Constraints on Jupiters from observations of Galactic Bulge microlensing events during 2000.
\thanks{Based on observations made with the Jacobus Kapteyn Telescope operated on the island of La Palma by the Isaac Newton Group in the Spanish Observatorio del Roque de los Muchachos of the Instituto de Astrofisica de Canarias.}}

\author[Tsapras, Horne, et al.]
{\parbox[t]{\textwidth}{ 
Yiannis~Tsapras$^1$, 
Keith~Horne$^{1,}$$^2$, Richard Carson$^1$,
Javier~M\'{e}ndez~Alvarez$^3$,\\
Dan~Batcheldor$^3$,
Alister W. Graham$^4$,
Philip A. James$^5$
Johan Knapen$^{3,}$$^6$,\\
Hannah~Quaintrell$^7$,
Ignacio~Gonzalez~Serrano$^3$, 
Peter~Sorensen$^3$,
Nick~Wooder$^8$\\
}\\
$^1$School of Physics and Astronomy, Univ. of St Andrews, Scotland KY16 9SS \\
$^2$Department of Astronomy, University of Texas, Austin TX 78712, USA\\
$^3$Isaac Newton Group of Telescopes (ING), La Palma, Spain\\
$^4$Instituto de Astrof\'{i}sica de Canarias,  E-38205 La Laguna, Tenerife, Spain\\
$^5$Astrophysics Research Institute, Liverpool John Moores University,
Birkenhead CH41 1LD, UK\\
$^6$ Department of Physical Sciences, University of Hertfordshire, Hatfield, Herts AL10 9AB, UK\\
$^7$Department of Physics, The Open University, Milton Keynes, UK\\
$^8$The Technology Partnership, Melbourn Science Park, Cambridge Road, Melbourn, S. Cambs. SG8 6EE, UK
}
\date{submitted Sep 2001}
\maketitle

\begin{abstract}
We present observations of 8 Galactic Bulge microlensing events taken with the 1.0m JKT  on La Palma during 2000 June and July. The JKT observing schedule was optimized using a prioritizing algorithm to automatically update the target list. For most of these events we have sampled the lightcurves at times where no information was available from the OGLE alert team. We assume a point-source point-lens (PSPL) model and perform a maximum likelihood fit to both our data and the OGLE data to constrain the event parameters of the fit. We then refit the data assuming a binary lens and proceed to calculate the probability of detecting planets with mass ratio $q=10^{-3}$. We have seen no clear signatures of planetary deviations on any of the 8 events and we quantify constraints on the presence of planetary companions to the lensing stars. For two well observed events, 2000BUL31 and 2000BUL33, our detection probabilities peak at $\sim$30\% and $\sim$20\% respectively for $q=10^{-3}$ and $a \sim R_{\mbox{E}}$ for a $\Delta\chi^2$ threshold value of 60.
\end{abstract}

\begin{keywords}
Stars: planetary systems, extra-solar planets, microlensing --
Techniques: photometric --
\end{keywords}

\section{INTRODUCTION}
Microlensing alters the path followed by the photons emitted by a background stellar source as they come near the influence of the gravitational field of a massive foreground object which acts as a lens. The separation of the images created by the lensing effect ($\sim 10^{-3}$arcsec) is too small to be resolved and only the combined flux is observed. The resulting lightcurve is symmetric in time with its maximum amplification at the time of closest approach between the projected position of the source on the lens plane and the lens itself. Its shape is well described by the formula:
\begin{equation}
A(u) = \frac{u^2 + 2}{u (u^2 + 4)^{1/2}}
\end{equation}
where $A(u)$ is the total amplification and $u$ is the angular separation of source and lens in units of the angular Einstein ring radius $\theta_{\mbox{E}}$ \cite{Pacz86}.

Even though most of the cases can be adequately described by this simple model, the shape of the lightcurve may not be symmetric and may be exhibiting significant deviations. These so-called anomalies of the lightcurve can be due to several factors and have been extensively examined in recent literature \cite{Dominik99b,Wozniak97,Buchalter97,Alcock95b,Gaudi97}. Arguably the most interesting of these are the anomalies which can be attributed to the binary nature of the lens. The possibility of the secondary component being an object of planetary characteristics has spawned dedicated observing campaigns to reveal their presence \cite{albrow98,rhie2000,bond01}.

\section{OUTLINE OF THE OBSERVATIONS}
From 2000 June 6 to July 17 we used the 1m Jacobus Kapteyn Telescope (JKT) on La Palma (Longitude: ${17^{\circ}52'41"}$ West, Latitude: ${28^{\circ}45'40"}$ North), Spain, to observe a number of microlensing events alerted by the OGLE (Optical Gravitational Lensing Experiment) team. A CCD SITE2 chip was mounted on the telescope with a gain value of 1.95 $e^{-}/$ADU and a readout noise of 11.7 $e^{-}$ in TURBO readout mode, linear to $\pm$1.1\% for exposures to 55k ADU. The pixel scale was 0.33 arcsec/pixel and the observations were taken using an $I$ filter at 820nm where the quantum efficiency of the chip is $\sim$60\%.

\begin{figure}
\centering
\begin{tabular}{c}
\psfig{file=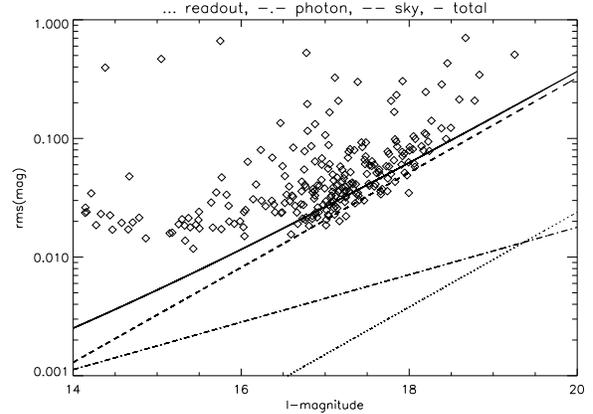,angle=0,width=8cm}
\end{tabular}
\caption{\small Rms values versus the corresponding Magnitude values of 17 measurements for all the stars found on the field of 2000BUL34. The dominant source of noise for $I \geq 17$mag is from the sky.}
\protect\label{plot1}
\end{figure}

Since the Galactic Bulge is a southern object, most of the observations were performed at high airmass ($\sim 2$) while our typical seeing ranged from 1 to 2 arcsec. This however does not prove to be as great a hindrance as one may expect since photometric accuracy down to $\sim 1-2$\% is still possible [\cite{Tsapras01} and Fig.~\ref{plot1}] for the brighter part of the lightcurves ($I$ $\leq 16 \ $mag). Exposure times were varied depending on the target current predicted brightness in the $I$-band so as to maximize the S/N while avoiding saturation. Two exposures were taken on each night for the brighter targets to permit discrimination of cosmic ray hits. For seven of the events followed we obtained data at times when OGLE lacked any. In section 3 we discuss our observing strategy. Section 4 of the paper deals with the photometric analysis of the data. We present the analysis of the lightcurves in section 5 together with the OGLE information \cite{OGLE}. Section 6 discusses the limits on planetary companions for the events. We summarize our results in section 7.

\section{OBSERVING STRATEGY}
Since our targets are only observable for about 3 hours per night from the JKT site during the summer, and since we were allocated only 2 hours per night for the observations, we had to fine tune the observations in order to maximize the number of events followed per night and also not to miss any events that were nearing maximum amplification.

For this purpose, we set up a daily auto-generated webpage to drive the observing schedule and which the observers could consult at any time. The events available on each night were listed, along with their RAs and Decs, predicted current $I$ magnitude, suggested exposure times, finder charts, current lightcurve fits and first estimations of the parameters. A (dimensionless) priority number was assigned to each event on a daily basis and the events available on each night were sorted and observed in order of importance. The priority number for each event was calculated using an empirical formula that places more weight on short, bright, high amplification events that are approaching their peaks.

Use of a prioritizing algorithm to optimize the follow-up observing of microlensing events will become more essential from 2002, when OGLE III comes online, since the number of microlensing events alerted is expected to rise from the current rate of $\sim$100 to perhaps 300-1000 events per year.

As automatic data reduction pipelines will soon be able to detect anomalous data points in real time ($\chi^2$ of single lens fit $\geq$ ${{\chi^2}_{\mbox{thr}}}$), subsequent observations can be immediately dedicated to verifying and recording the shape of the lightcurve anomaly.

\section{PHOTOMETRIC ANALYSIS}
The photometric analysis was performed using the IRAF/DAOPHOT package \cite{Davis94} in a semi-automated pipeline. For further processing and lightcurve fitting we used separate programs developed in IDL (Interactive Data Language).

Standard pre-processing, which involves subtraction of the bias level and division by the master flat field, is first applied to the CCD frames. Once this is complete, we register the frames to the nearest pixel using FIGARO so that the star positions correspond to the same pixel areas on each frame. We then crop the frames to 300 $\times$ 300 pixels ($99''\times 99''$), centered on the target, to reduce processing time since we are mainly interested in the target star and a certain number of nearby reference stars to determine the frame-to-frame magnitude corrections. The initial average sky value and its variance for each frame, which are needed to drive DAOPHOT, are determined by picking a star-free region and calculating the sky background photon statistics. The sky value is then recalculated within the routine.
\begin{figure*}
\def\subfigtopskip{4pt}
\def\subfigbottomskip{8pt}
\def\subfigcapskip{4pt}
\centering
\begin{tabular}{cc}

\subfigure[2000BUL26]{\label{fig:bul26}
\psfig{file=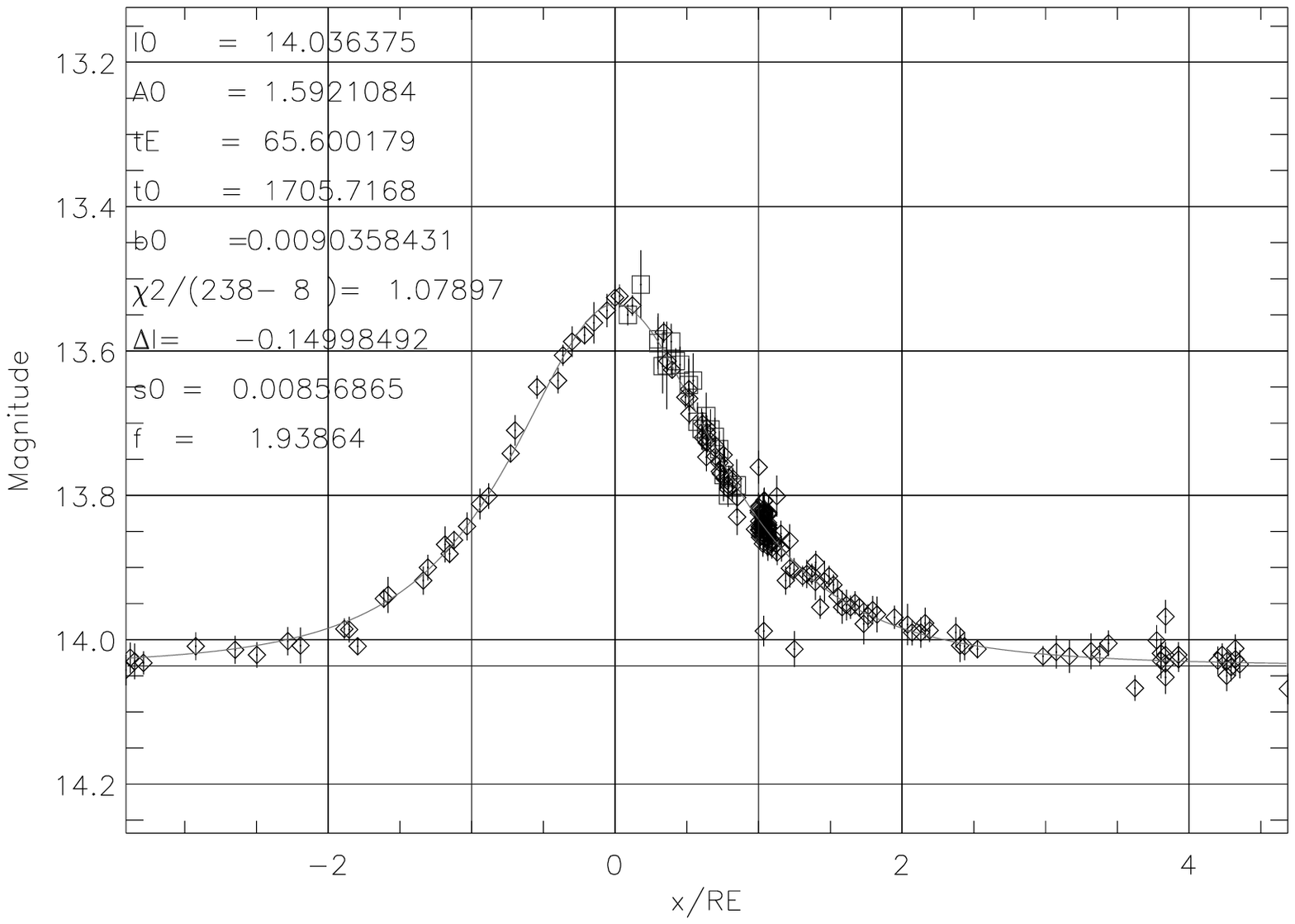,angle=0.0,width=8cm}}
&
\subfigure[2000BUL29]{\label{fig:bul29}
\psfig{file=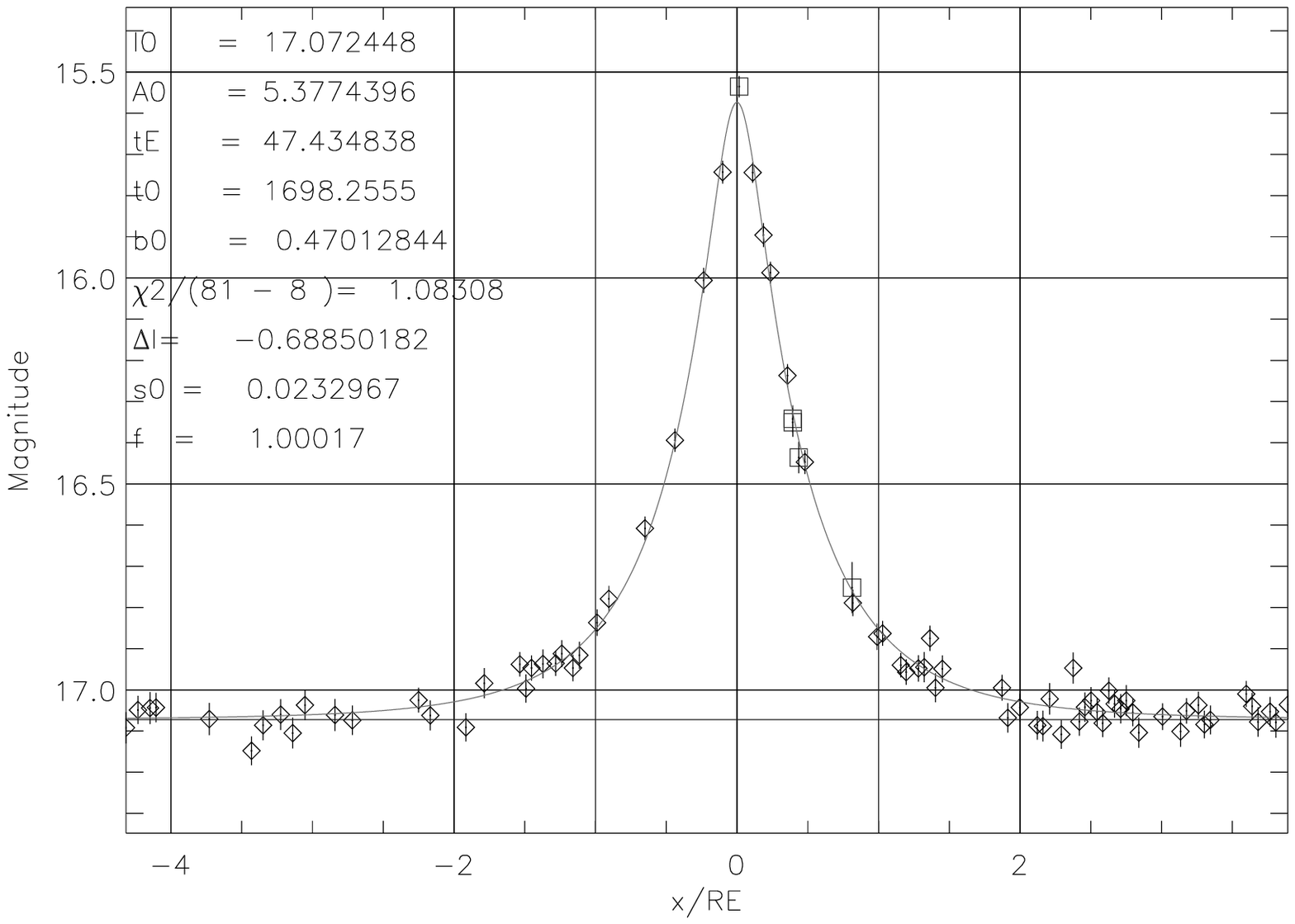,angle=0.0,width=8cm}}\\

\subfigure[2000BUL31]{\label{fig:bul31}
\psfig{file=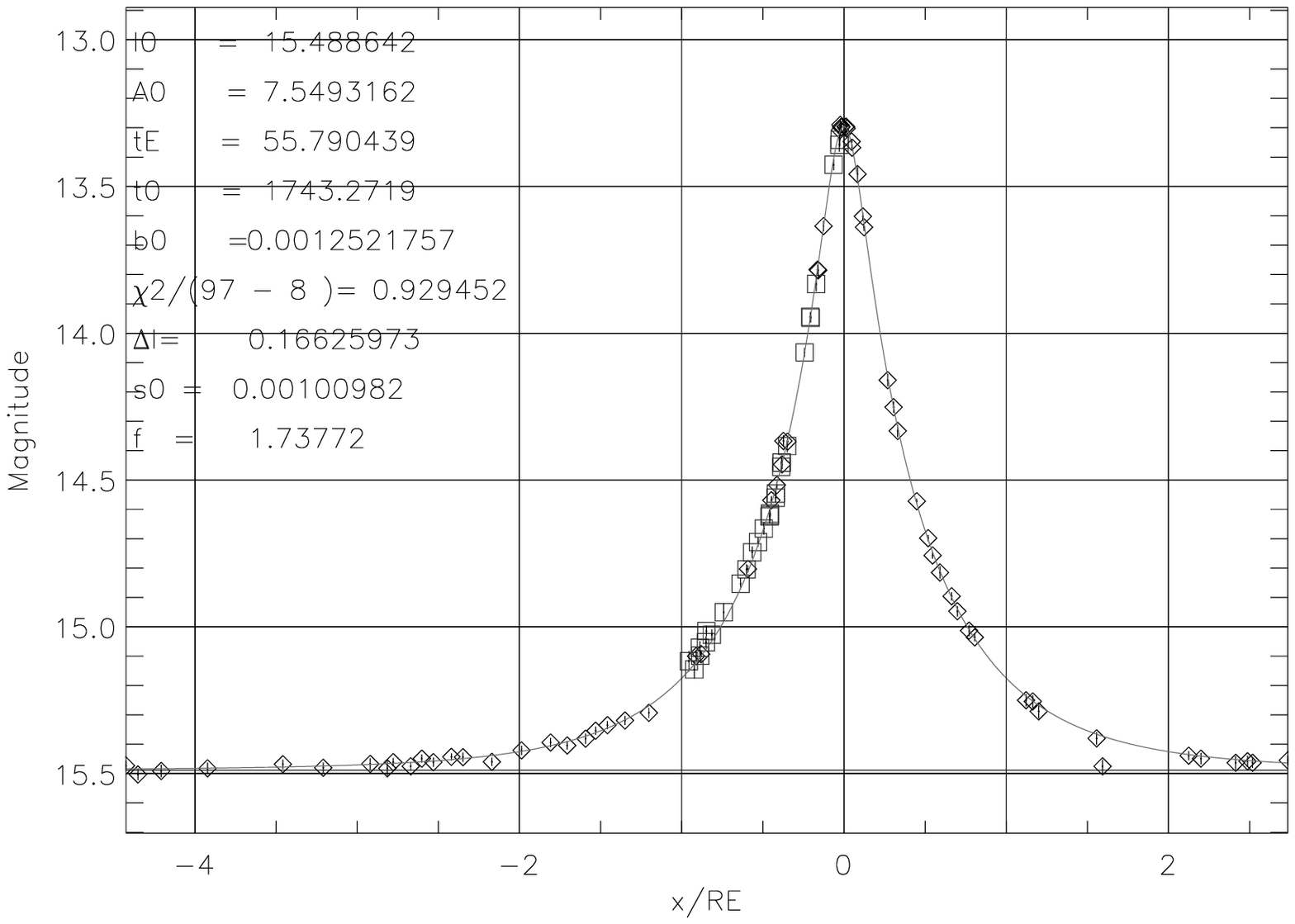,angle=0.0,width=8cm}}
&
\subfigure[2000BUL33]{\label{fig:bul33}
\psfig{file=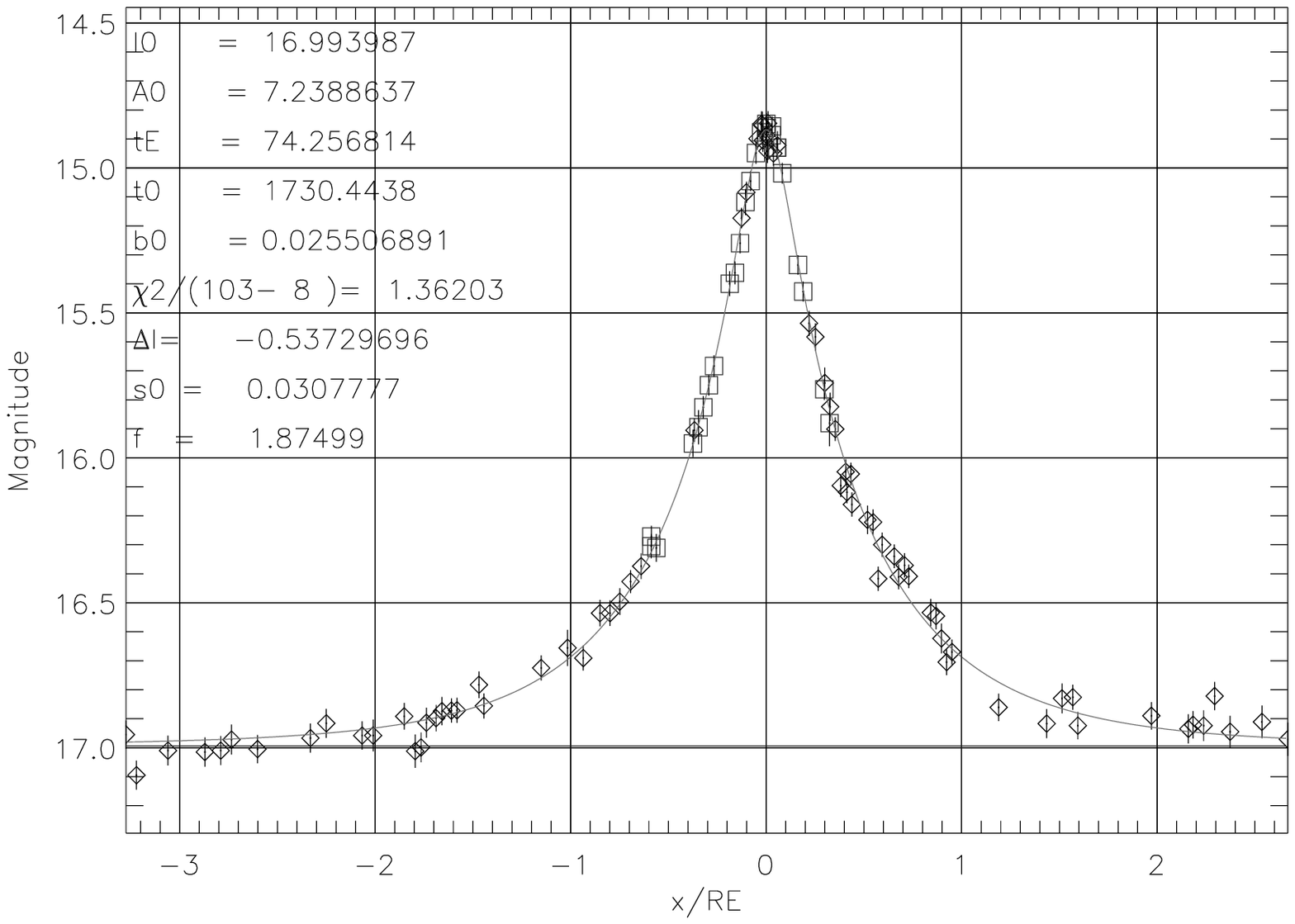,angle=0.0,width=8cm}}\\

\end{tabular}
\caption{$\chi^2$ minimization fits to the combined data. The squares indicate JKT data and the diamonds OGLE data.}
\protect\label{fig:magplots1}
\end{figure*}

The next step involves identifying all the stars on a good seeing frame by using the DAOFIND task. A number ($\sim$ 10-15) of bright, uncrowded stars are manually selected as point-spread function (PSF) stars which are used to create the PSF for the frame. The profile uses the `penny2' analytic function which comprises of a Gaussian core with Lorentzian wings. 
\begin{table*}
\centering
\caption{Fitted parameters for the events observed and reduced $\chi^2$ values.}
\protect\label{tab:observations}
\vspace{5mm}
\begin{tabular}{lcccccccccccccl}
\hline
EVENT  &  $N_{\mbox{JKT}}$ & $N_{\mbox{OGLE}}$ & $I_{\mbox{OGLE}}$ & $\Delta I$  & $A_{0}$ & $t_{\mbox{E}}$(d) & $t_{0}$(HJD 245+) & $\frac{\chi^2}{(N-6)}$ & $f$ & $\sigma_{0}$(mag) & $\frac{\chi^2}{(N-8)}$ & $b_0$ \\
\hline
BUL26 & 21 & 217 & 14.04&  $-0.150$ & 1.59 & 65.60 & 1705.72 & 5.29 & 1.939 & 0.009 & 1.08 & 0.009\\
BUL29 & 5  & 76 & 17.07 &  $-0.689$ & 3.82 & 47.43 & 1698.26 & 2.21 & 1.000 & 0.023 & 1.08 & 0.470\\
BUL31 & 29 & 68 & 15.49 &  $+0.166$ & 7.55 & 55.79 & 1743.27 & 3.28 & 1.738 & 0.001 & 0.93 & 0.001\\
BUL33 & 26 & 77 & 16.99 &  $-0.537$ & 7.24 & 74.26 & 1730.44 & 9.47 & 1.875 & 0.031 & 1.36 & 0.026\\
BUL34 & 17 & 72 & 17.73 &  $+0.287$ & 5.23 & 27.04 & 1710.71 & 2.25 & 1.441 & 0.002 & 1.11 & 0.030\\
BUL36 & 18 & 69 & 16.06 &  $-0.143$ & 1.25 & 49.03 & 1711.66 & 2.67 & 1.628 & 0.001 & 1.04 & 0.900\\
BUL37 & 16 & 48 & 17.62 &  $+0.712$ & 2.24 & 33.08 & 1713.88 & 1.98 & 1.227 & 0.012 & 1.25 & 0.900\\
BUL39 & 13 & 61 & 17.11 &  $+0.712$ & 1.60 & 34.87 & 1719.34 & 3.25 & 1.679 & 0.004 & 1.19 & 0.888\\
\hline
\end{tabular}
\end{table*}

We then run ALLSTAR on our frames to measure all the stars present in the list. The magnitudes of the PSF stars are used to calculate the average magnitude offset of each star from its weighted average over time. The average of the offsets is the zero-point value $z_{i}$. This frame-dependent zero-point is then applied to the magnitude values of all measured stars on every frame. Analytically, the weighted average value of star $m_{j}$, for each frame $i$, is calculated using:
\begin{equation}
<m_{j}>=\displaystyle\frac{\displaystyle\sum_{i}{m_{ji} \ w_{ji}}}{\displaystyle\sum_{i}{w_{ji}}},
\end{equation}
where the weights used are inverse variance weights $w_{ji}=1/{\sigma^2}_{ji}$, 
$m_{ji}$ is the magnitude value of star $j$ on frame $i$ and the zero point for frame $i$ is:
\begin{equation}
z_{i}=\displaystyle\frac{\displaystyle\sum_{j}{(m_{ji} - <m_{j}>) \ w_{ji}}}{{\displaystyle\sum_{j}{w_{ji}}}}.
\end{equation}
Instrumental magnitudes $I_{ji}=m_{ji}-z_{i}$ rescaled in this fashion were then plotted vs time $t$ to produce the lightcurves.

The typical accuracy obtained by our method is illustrated on figure~\ref{plot1} which shows, for all stars found on the frames of the field of event 2000BUL34, the RMS scatter about the weighted average of 17 calibrated magnitude values. 
\begin{figure*}
\def\subfigtopskip{4pt}
\def\subfigbottomskip{8pt}
\def\subfigcapskip{4pt}
\centering
\begin{tabular}{cc}

\subfigure[2000BUL26]{\label{fig:bul26r}
\psfig{file=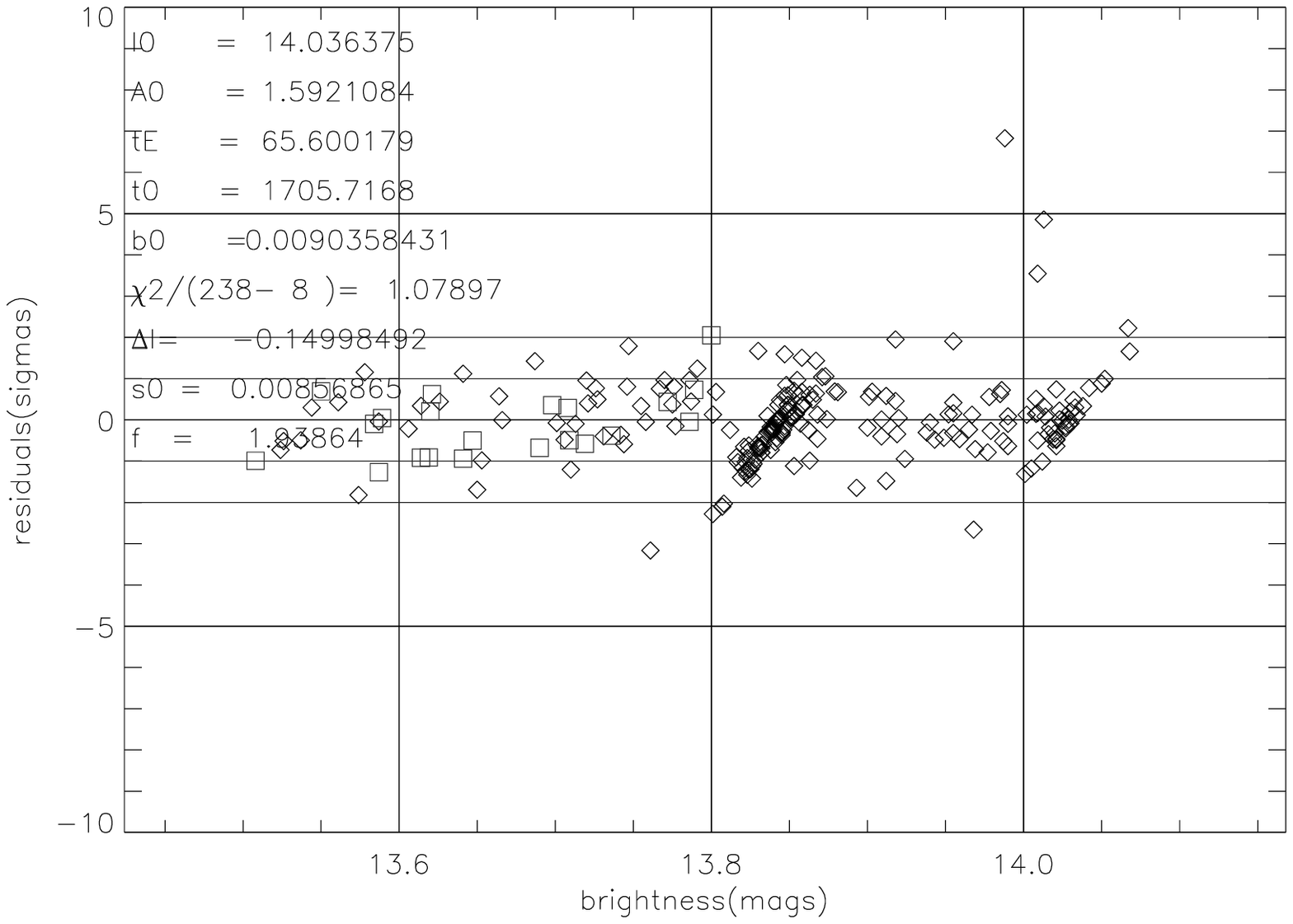,angle=0.0,width=8cm}}
&
\subfigure[2000BUL29]{\label{fig:bul29r}
\psfig{file=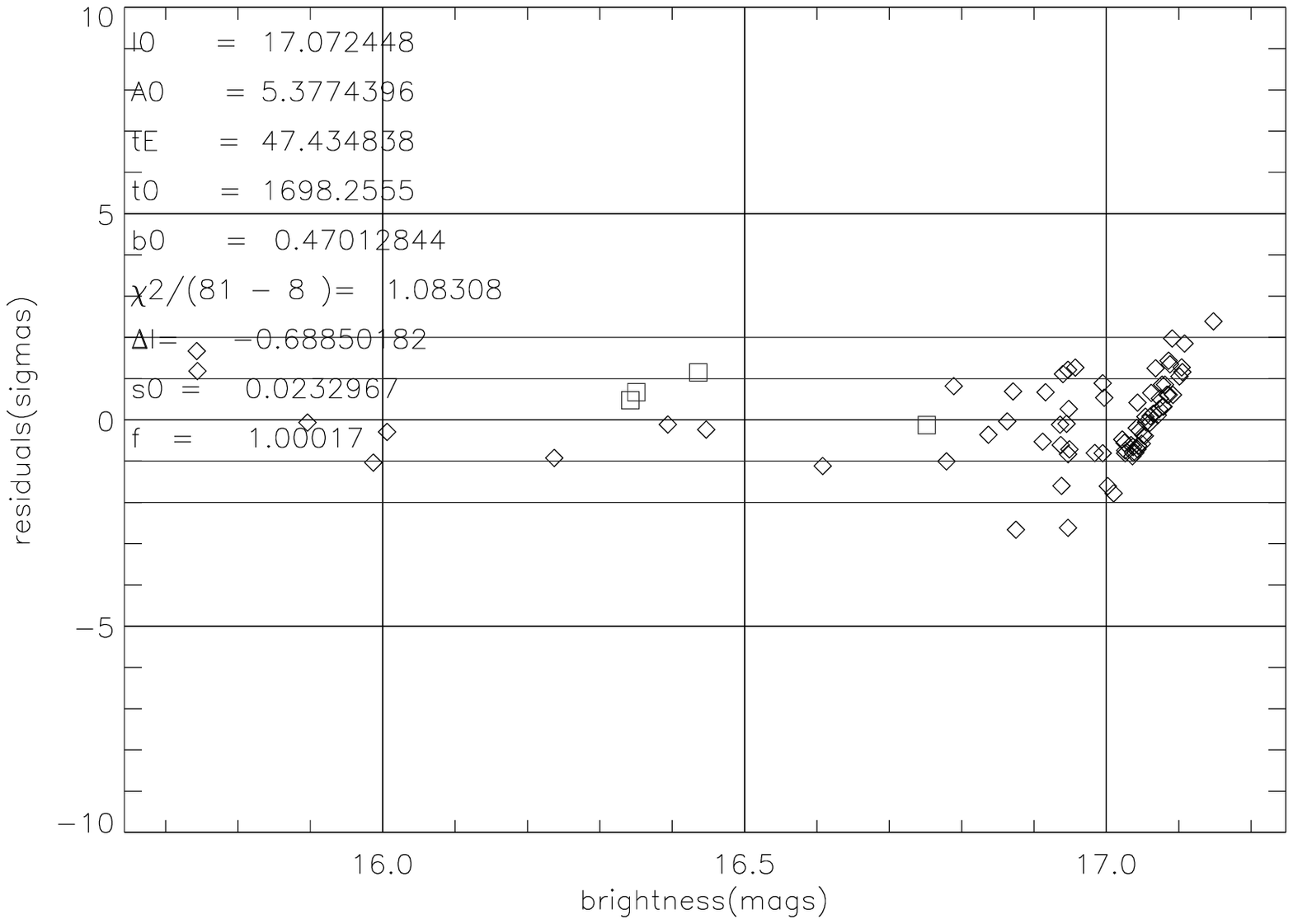,angle=0.0,width=8cm}}\\

\subfigure[2000BUL31]{\label{fig:bul31r}
\psfig{file=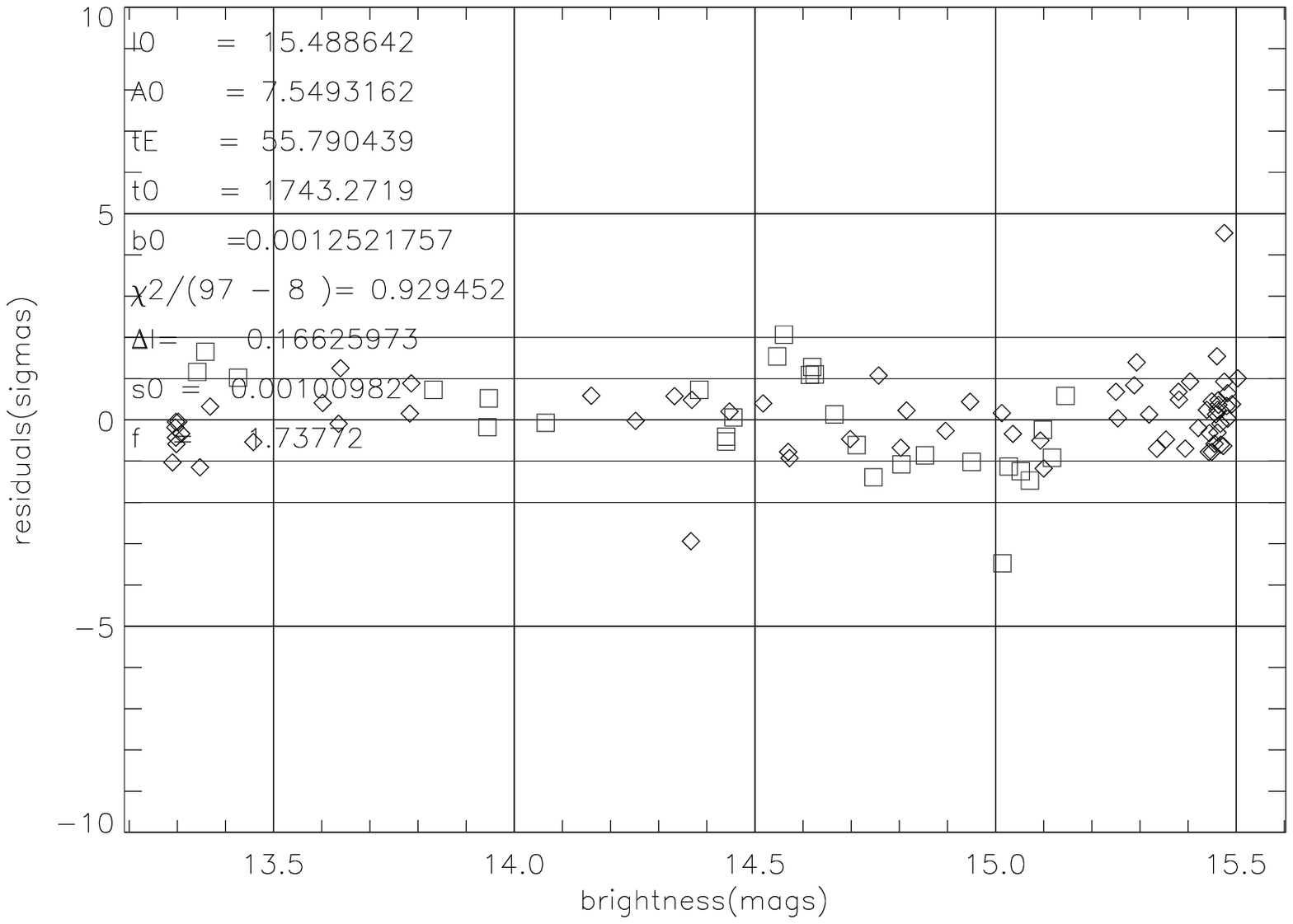,angle=0.0,width=8cm}}
&
\subfigure[2000BUL33]{\label{fig:bul33r}
\psfig{file=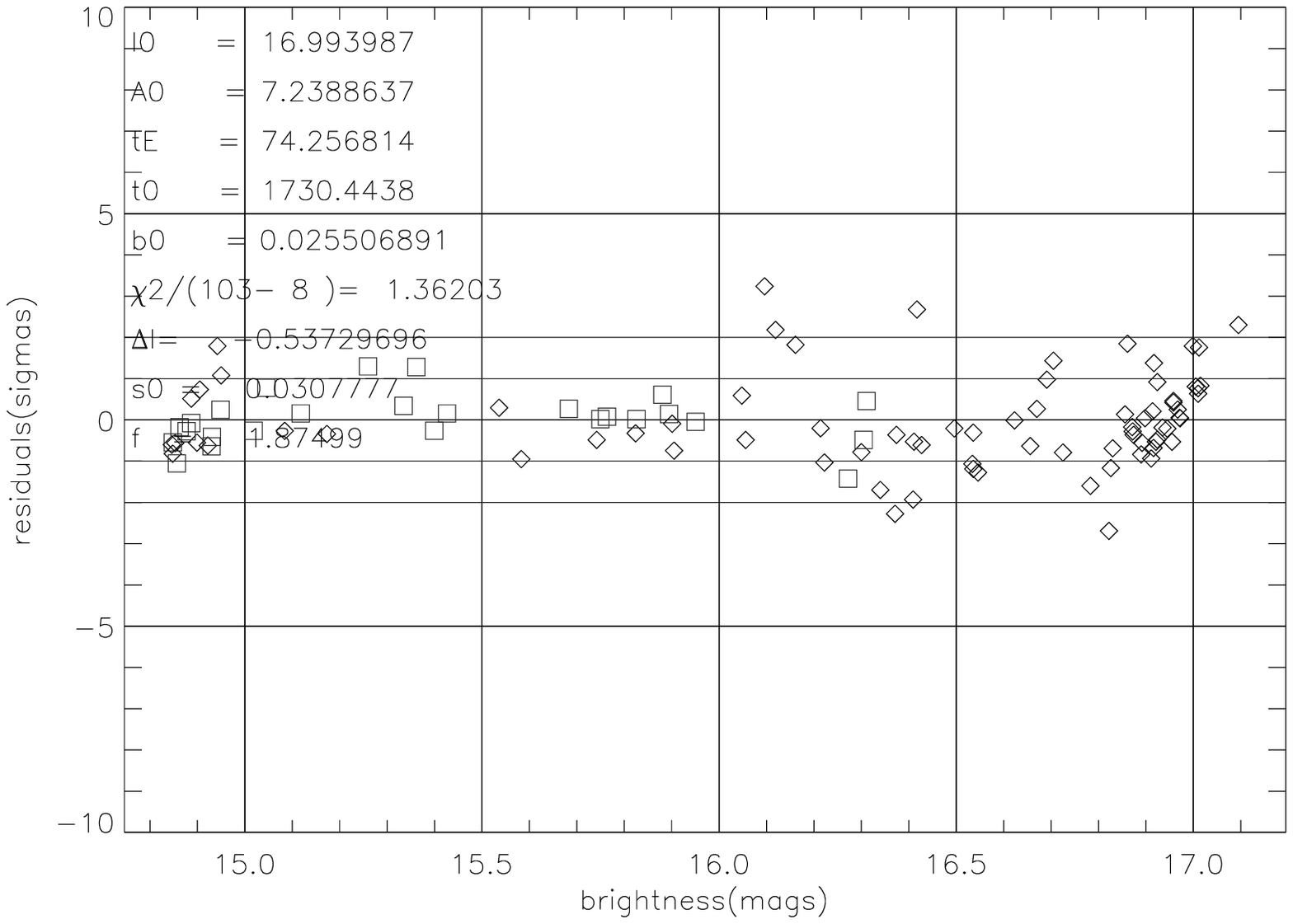,angle=0.0,width=8cm}}\\

\end{tabular}
\caption{Fit residuals of the combined data. The squares indicate JKT data and the diamonds OGLE data.}
\protect\label{fig:resplots1}
\end{figure*}

\begin{figure*}
\def\subfigtopskip{4pt}
\def\subfigbottomskip{8pt}
\def\subfigcapskip{4pt}
\centering
\begin{tabular}{cc}

\subfigure[2000BUL34]{\label{fig:bul34}
\psfig{file=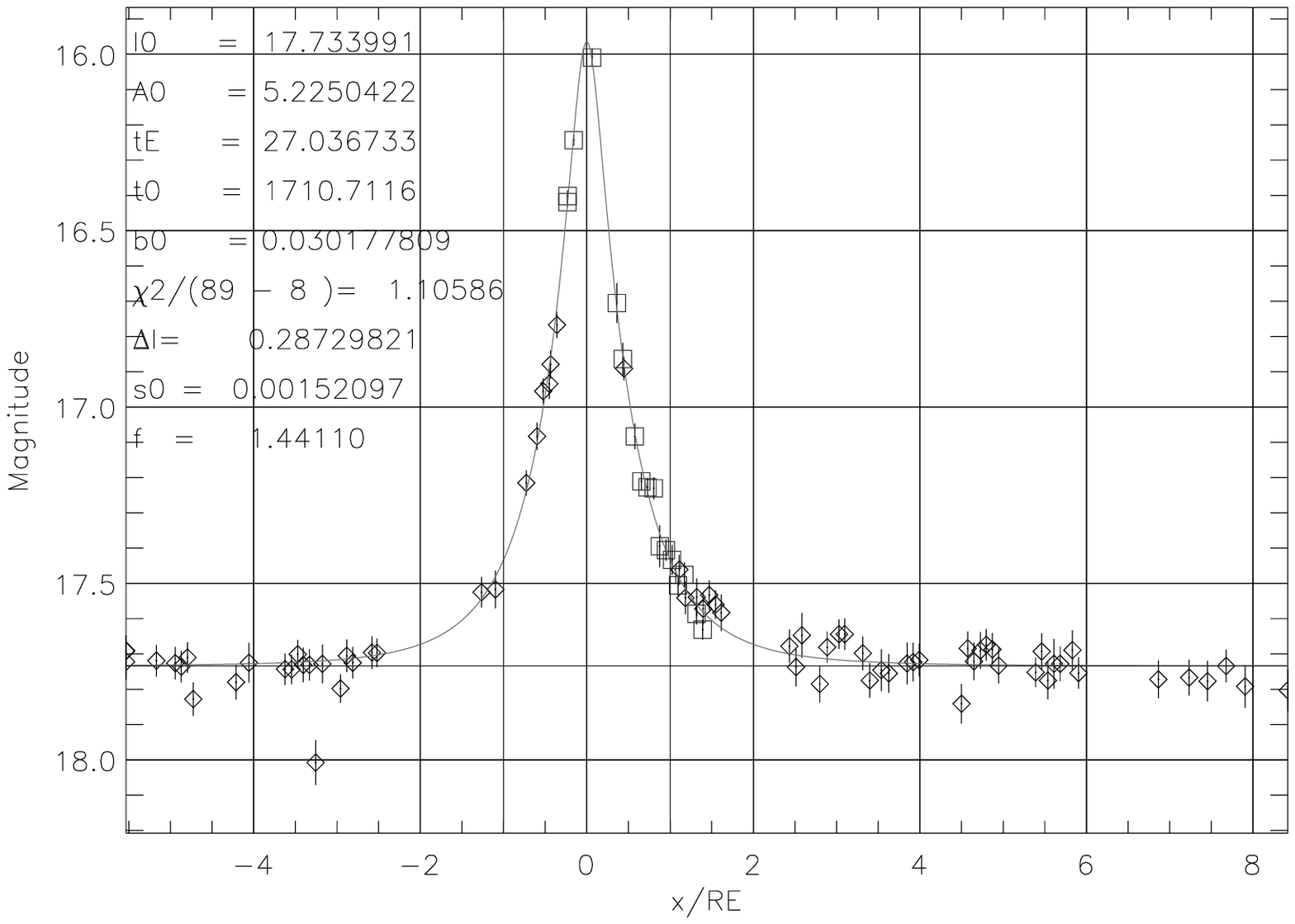,angle=0.0,width=8cm}} 
&
\subfigure[2000BUL36]{\label{fig:bul36}
\psfig{file=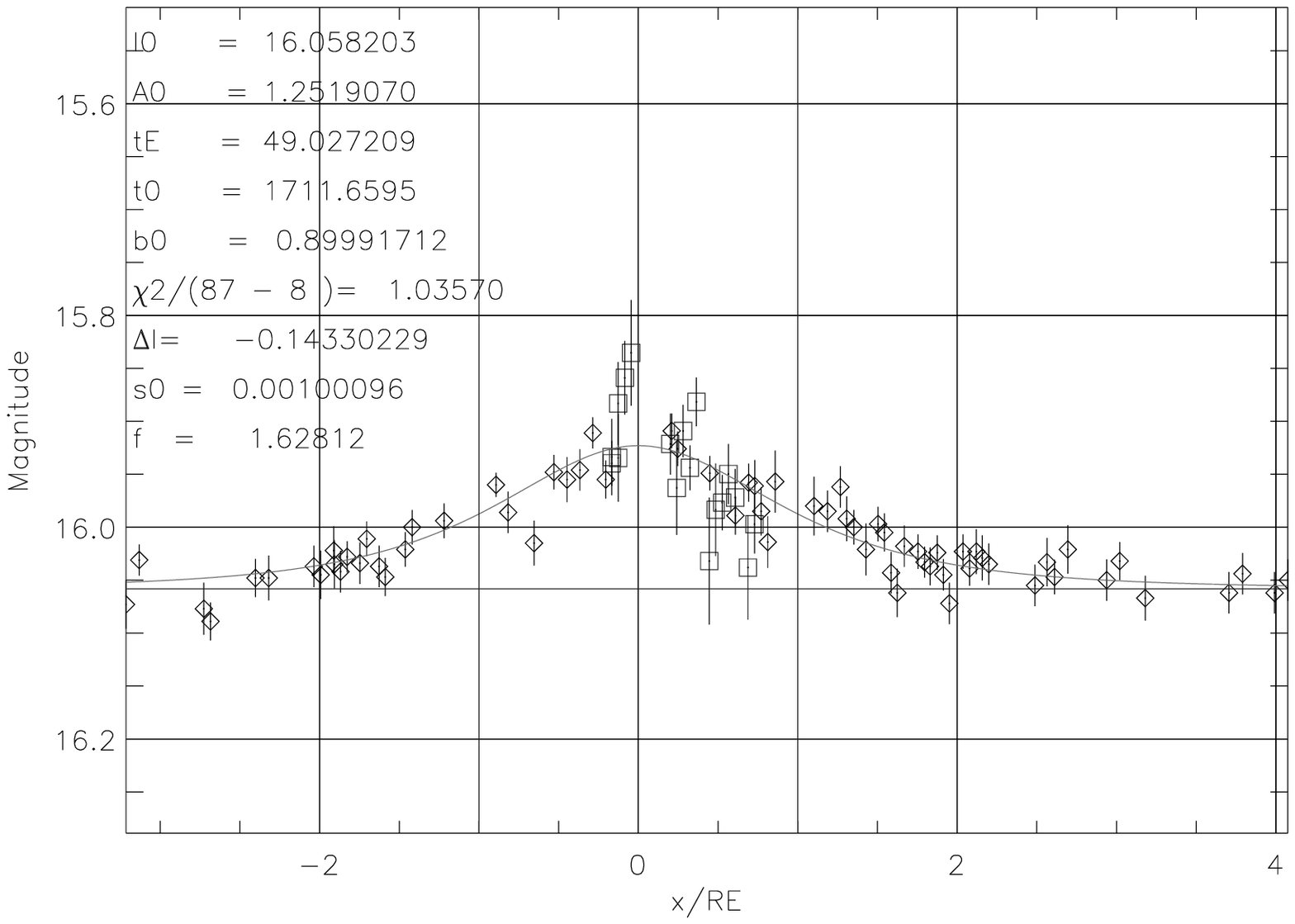,angle=0.0,width=8cm}} \\

\subfigure[2000BUL37]{\label{fig:bul37}
\psfig{file=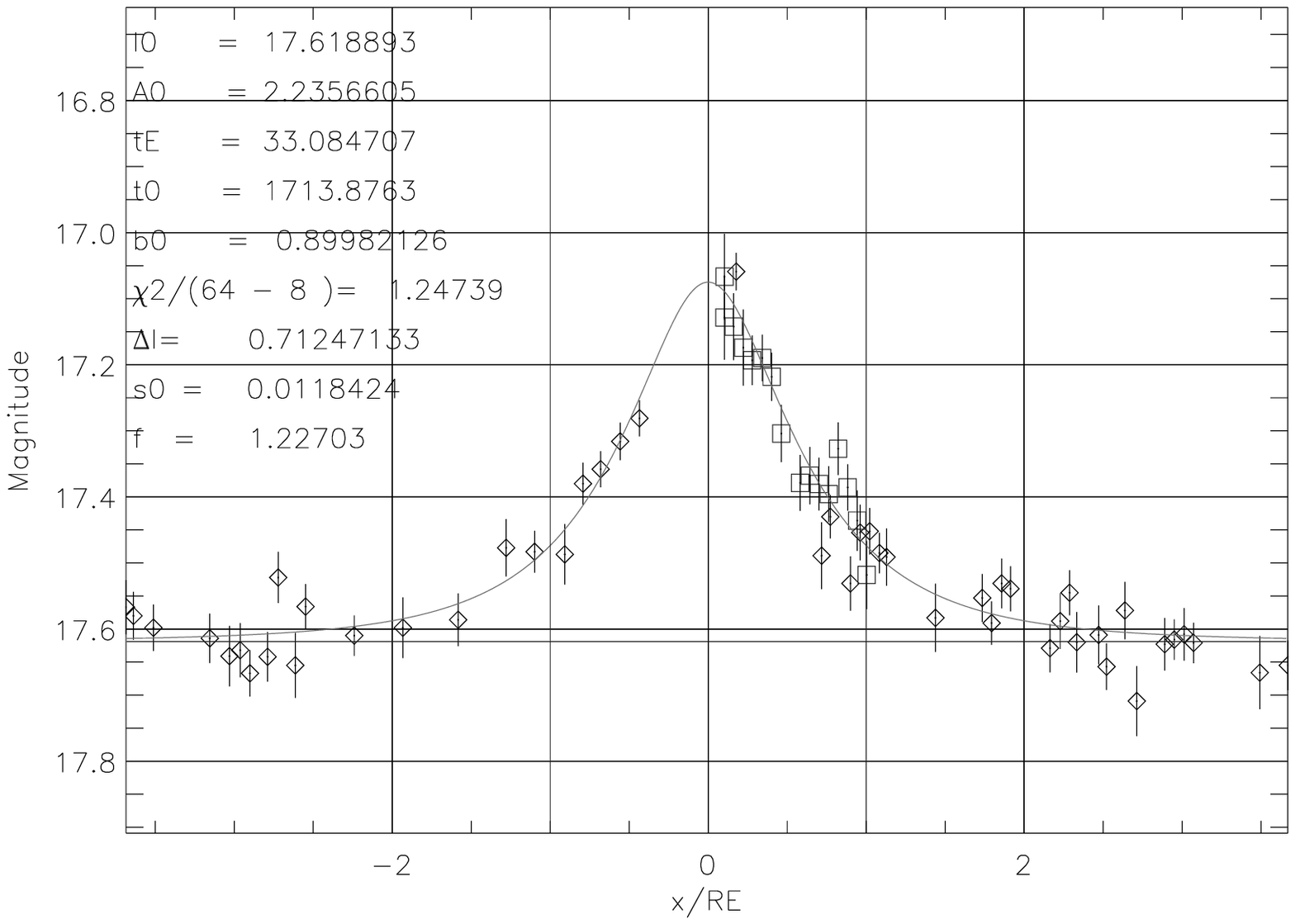,angle=0.0,width=8cm}}
&
\subfigure[2000BUL39]{\label{fig:bul38}
\psfig{file=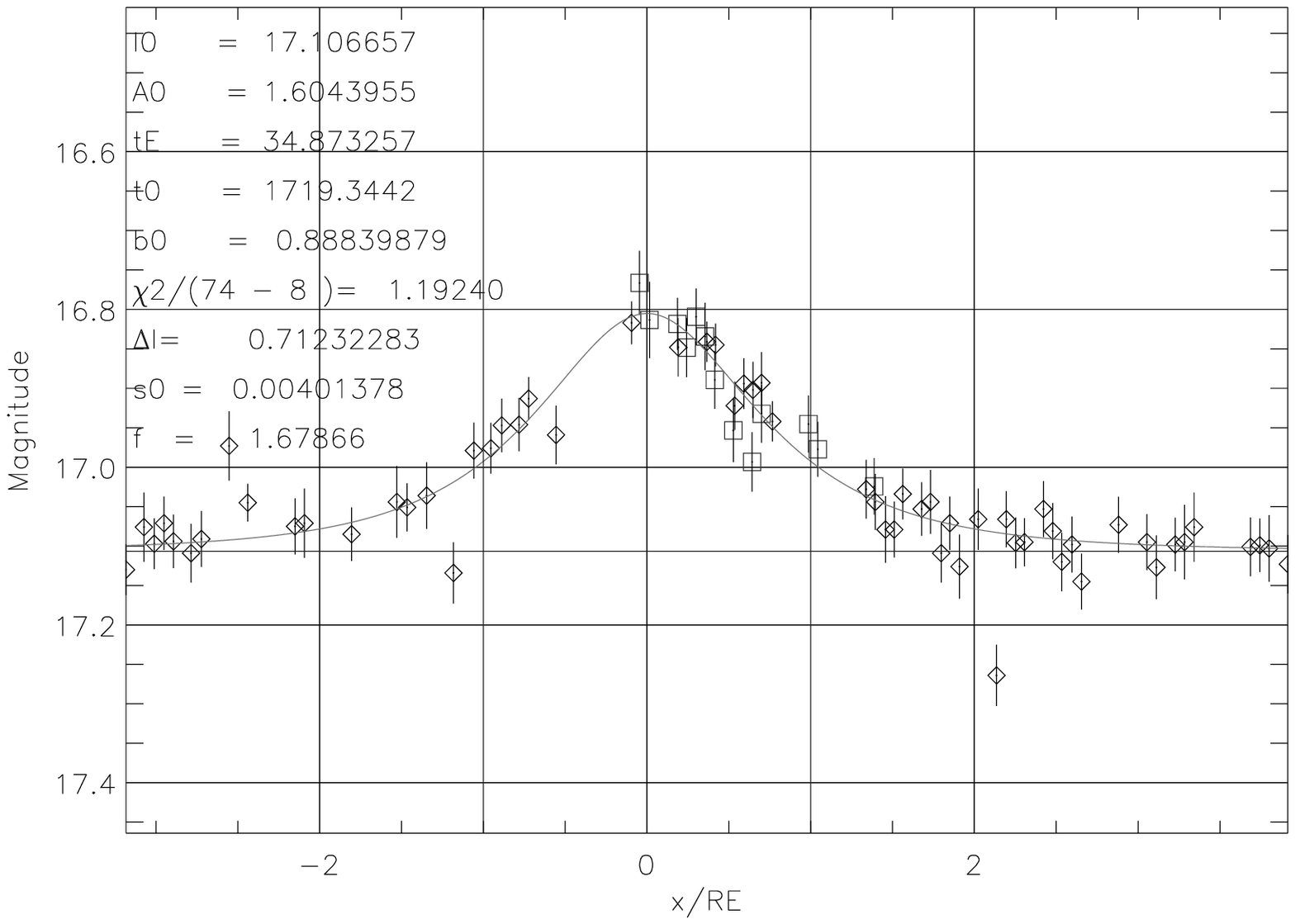,angle=0.0,width=8cm}} \\

\end{tabular}
\caption{$\chi^2$ minimization fits to the combined data. The squares indicate JKT data and the diamonds OGLE data.}
\protect\label{fig:magplots2}
\end{figure*}

\begin{figure*}
\def\subfigtopskip{4pt}
\def\subfigbottomskip{8pt}
\def\subfigcapskip{4pt}
\centering
\begin{tabular}{cc}

\subfigure[2000BUL34]{\label{fig:bul34r}
\psfig{file=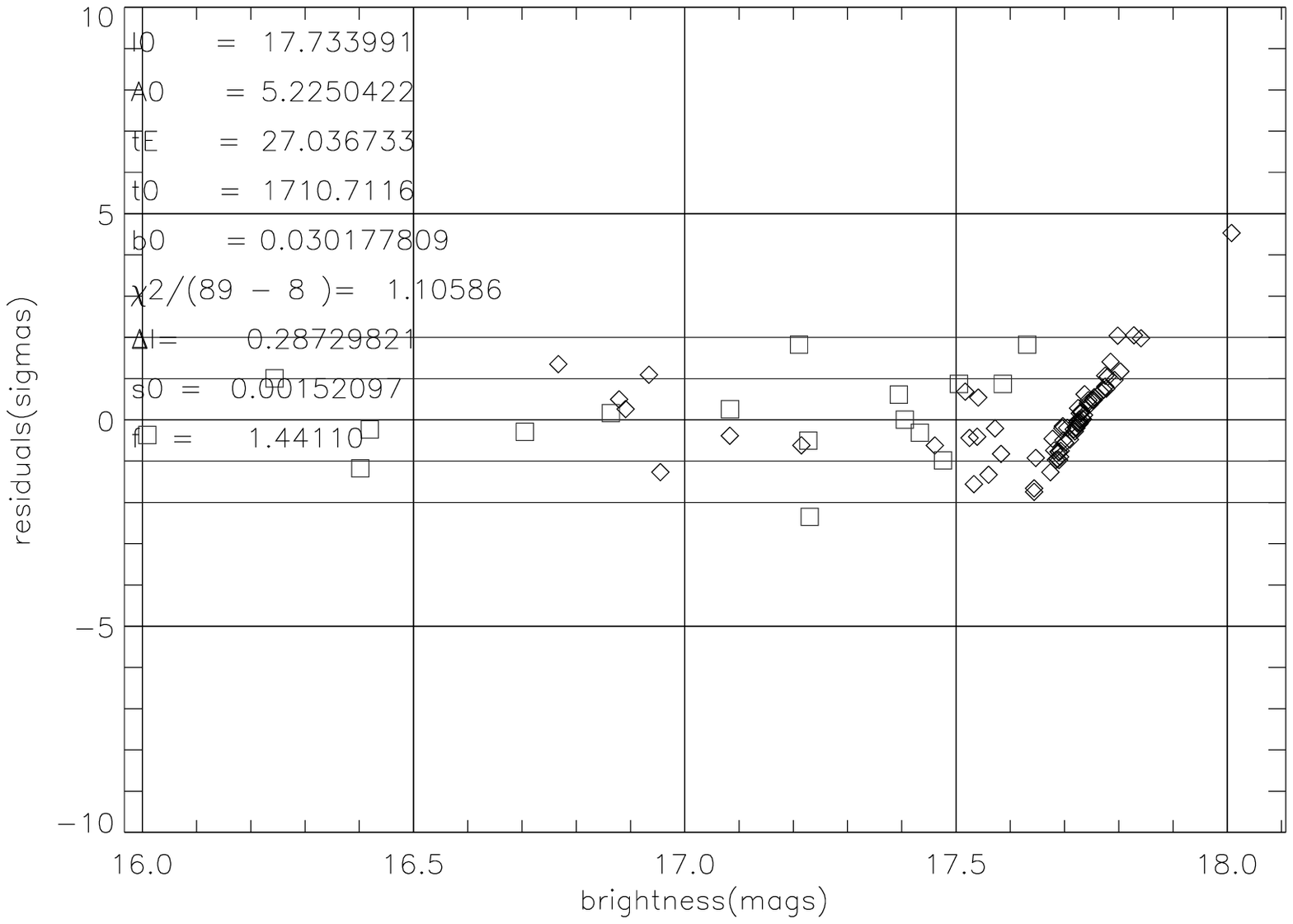,angle=0.0,width=8cm}} 
&
\subfigure[2000BUL36]{\label{fig:bul36r}
\psfig{file=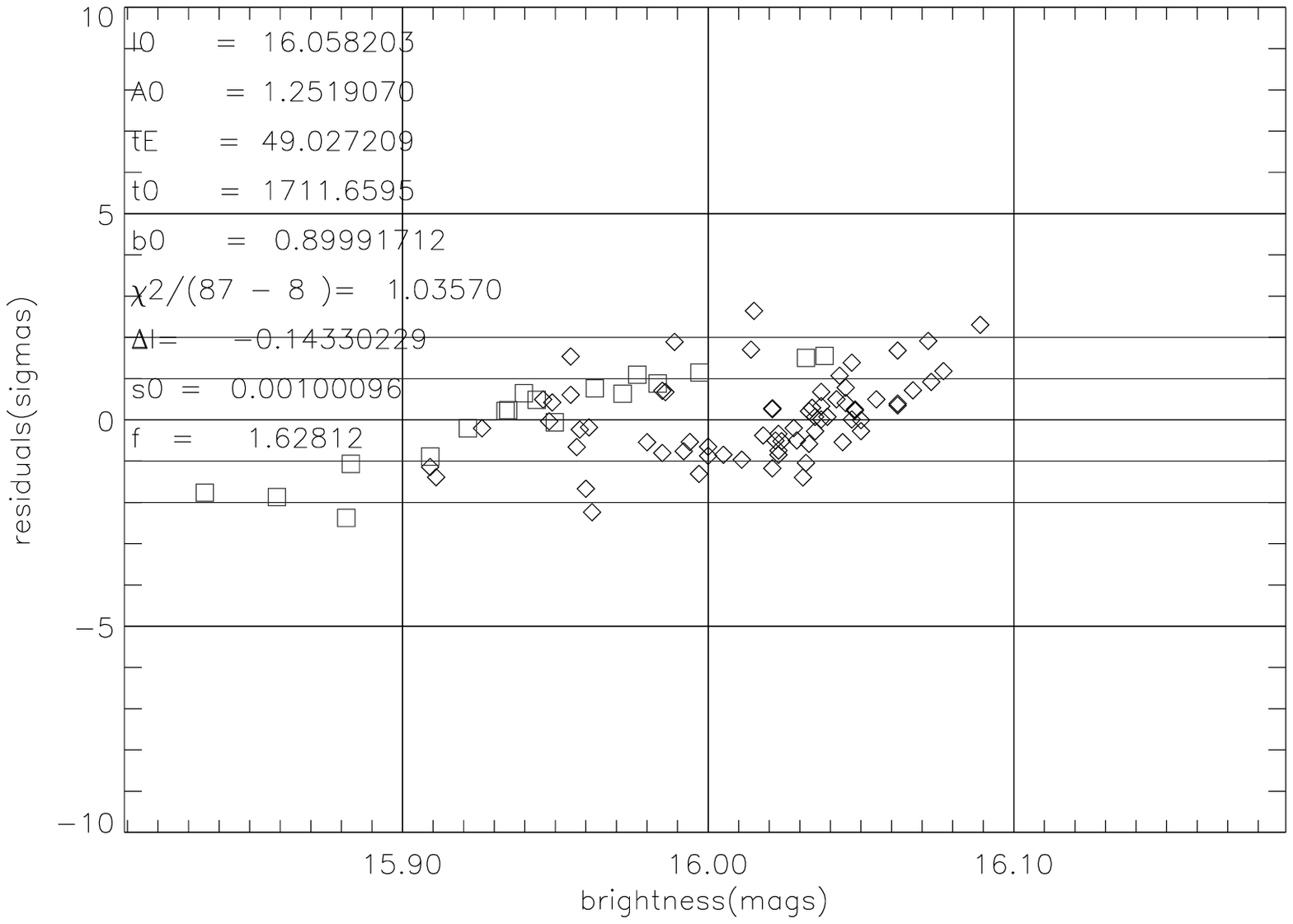,angle=0.0,width=8cm}} \\

\subfigure[2000BUL37]{\label{fig:bul37r}
\psfig{file=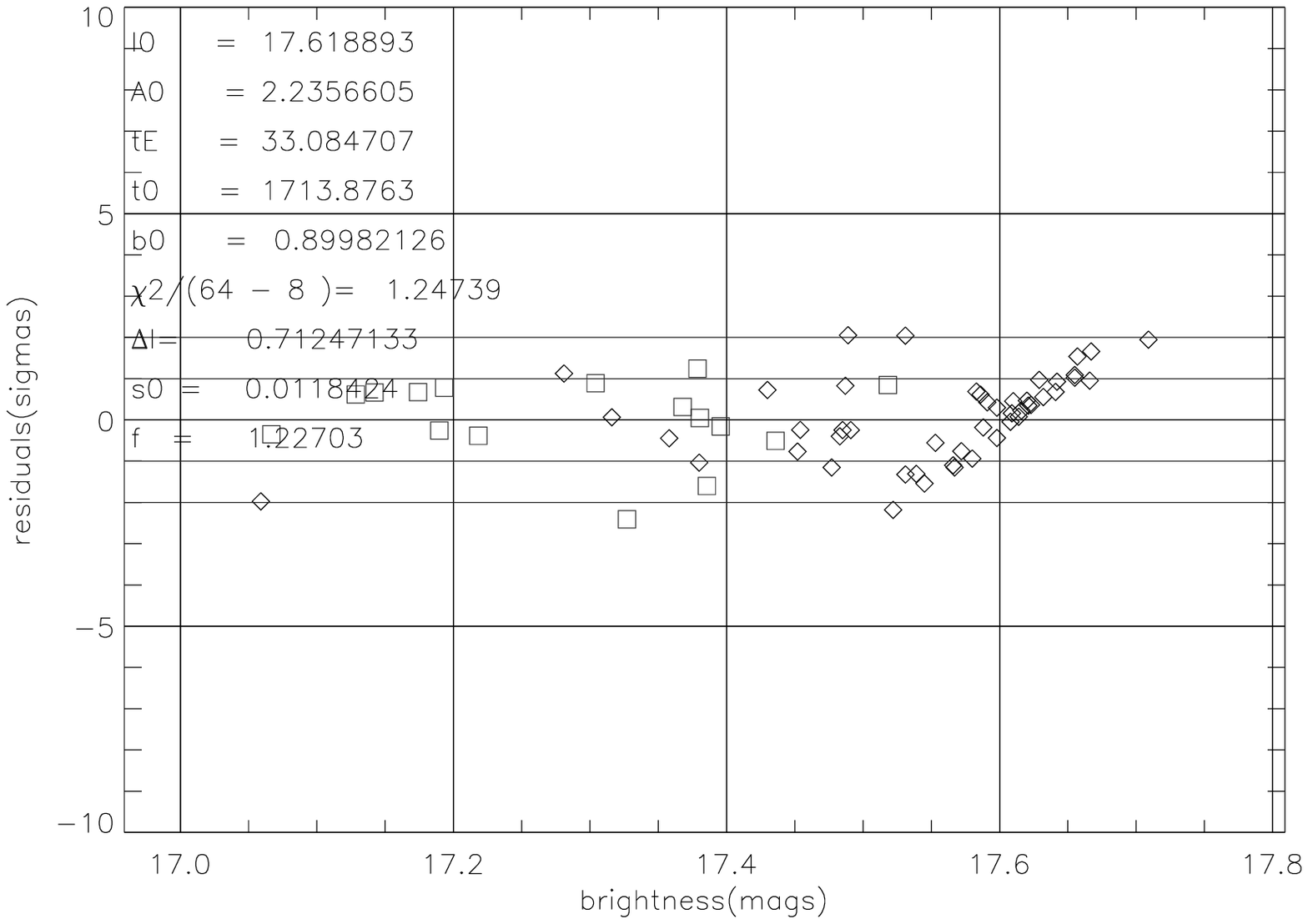,angle=0.0,width=8cm}}
&
\subfigure[2000BUL39]{\label{fig:bul38r}
\psfig{file=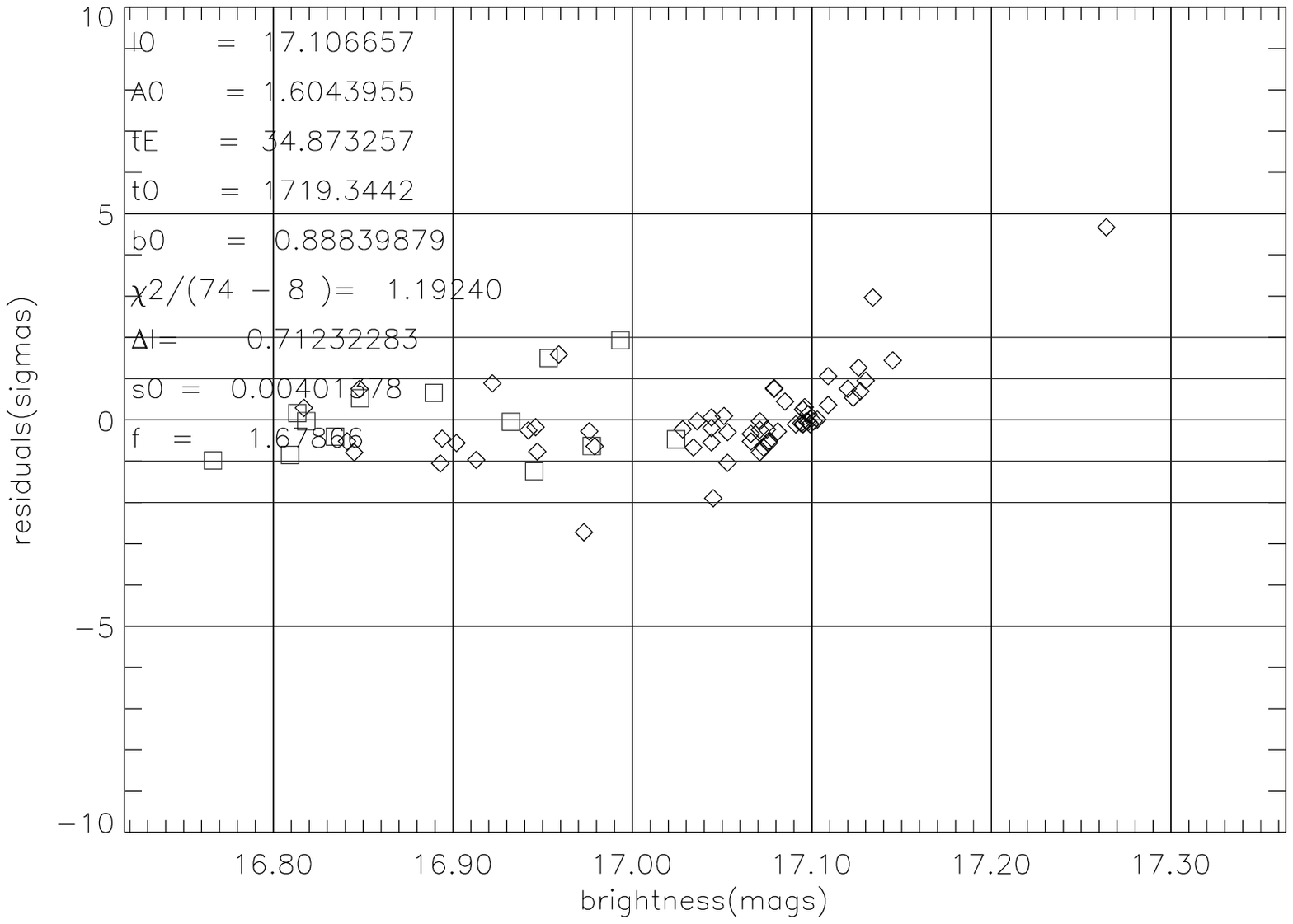,angle=0.0,width=8cm}} \\

\end{tabular}
\caption{Fit residuals of the combined data. The squares indicate JKT data and the diamonds OGLE data.}
\protect\label{fig:resplots2}
\end{figure*}

\begin{figure*}
\def\subfigtopskip{0pt}
\def\subfigbottomskip{0pt}
\def\subfigcapskip{0pt}
\centering
\begin{tabular}{ccc}
\subfigure[2000BUL26]{\label{fig:bul26p}
\psfig{file=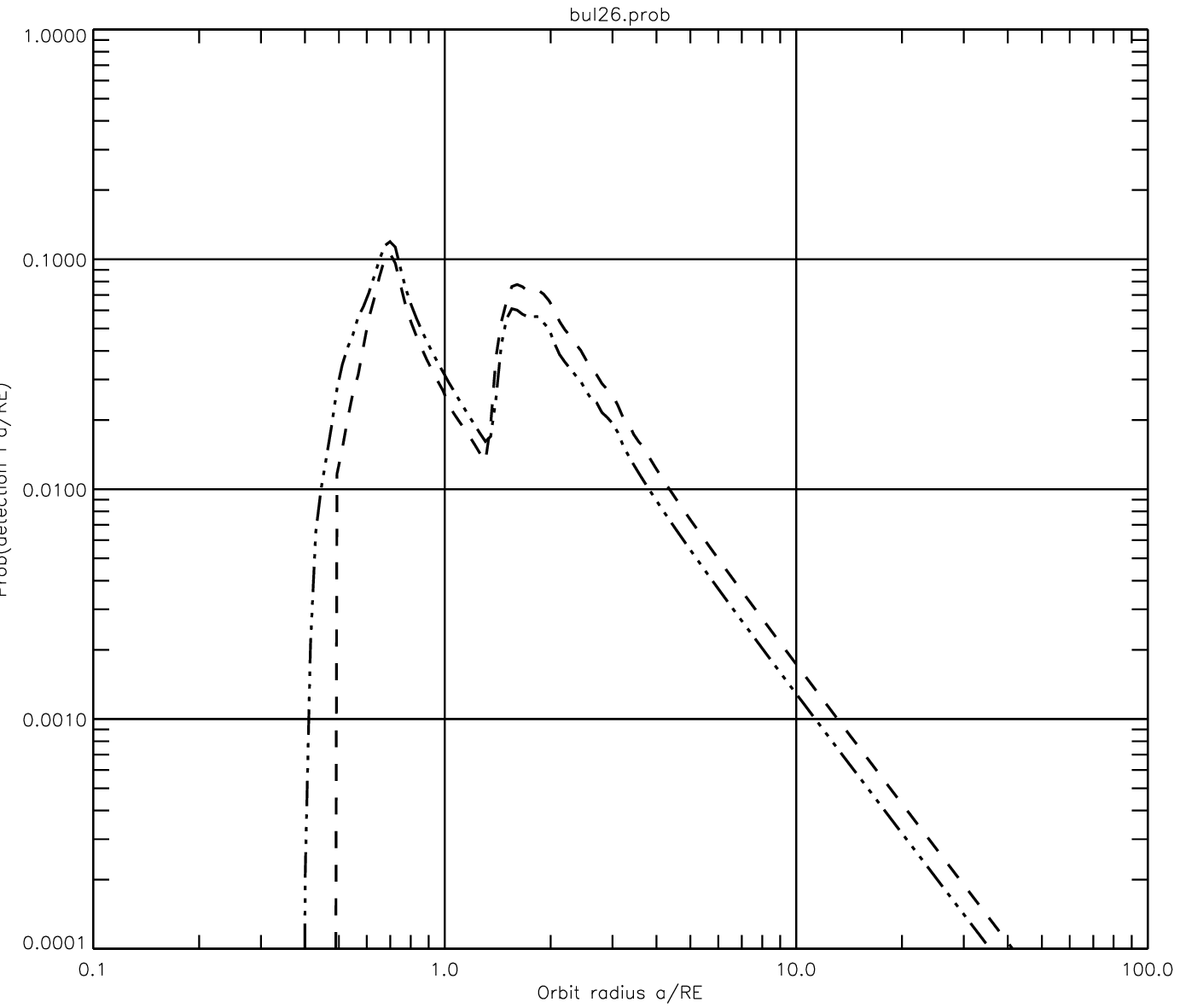,angle=0.0,width=5cm}}
&
\subfigure[2000BUL29]{\label{fig:bul29p}
\psfig{file=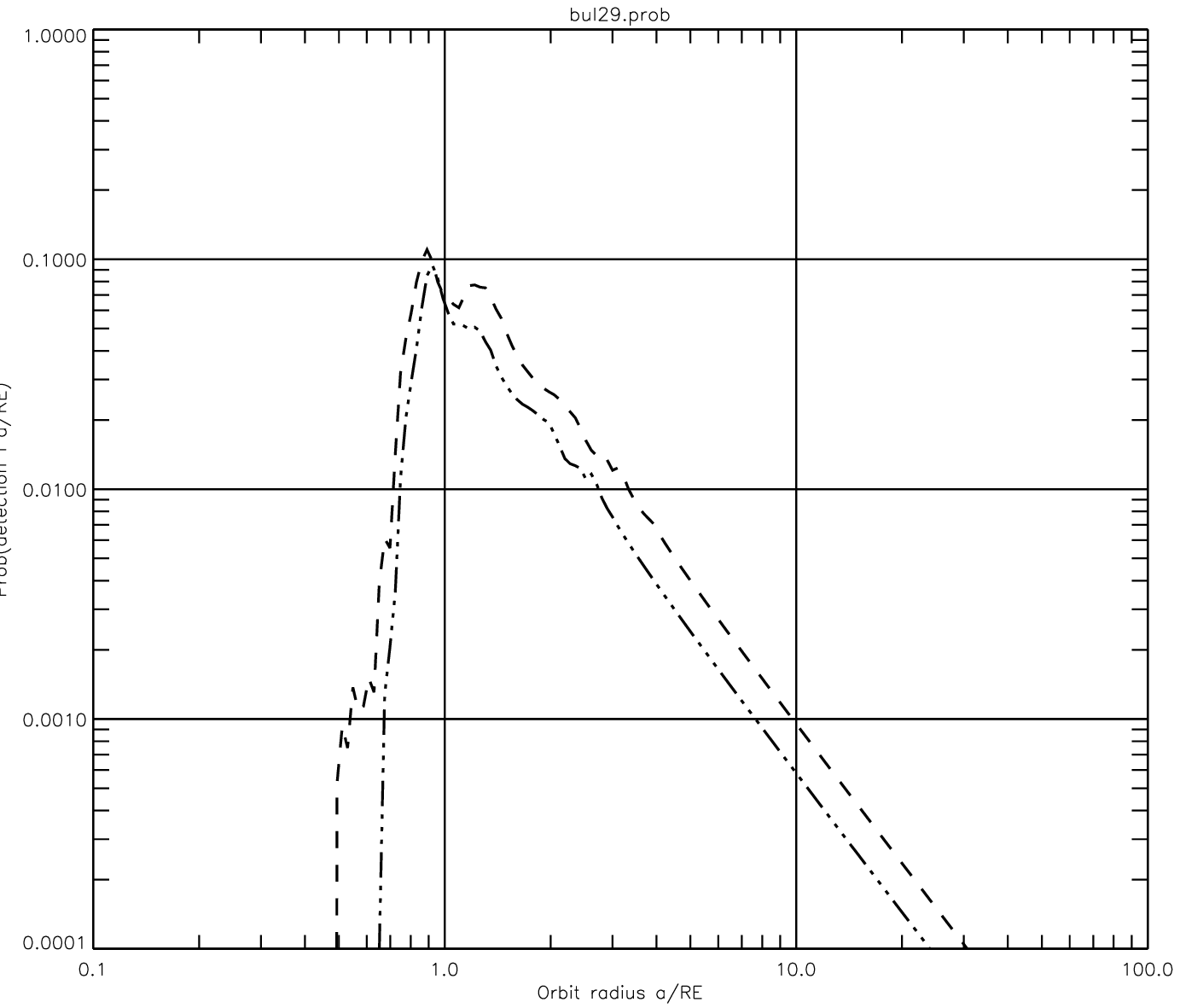,angle=0.0,width=5cm}}
&
\subfigure[2000BUL31]{\label{fig:bul31p}
\psfig{file=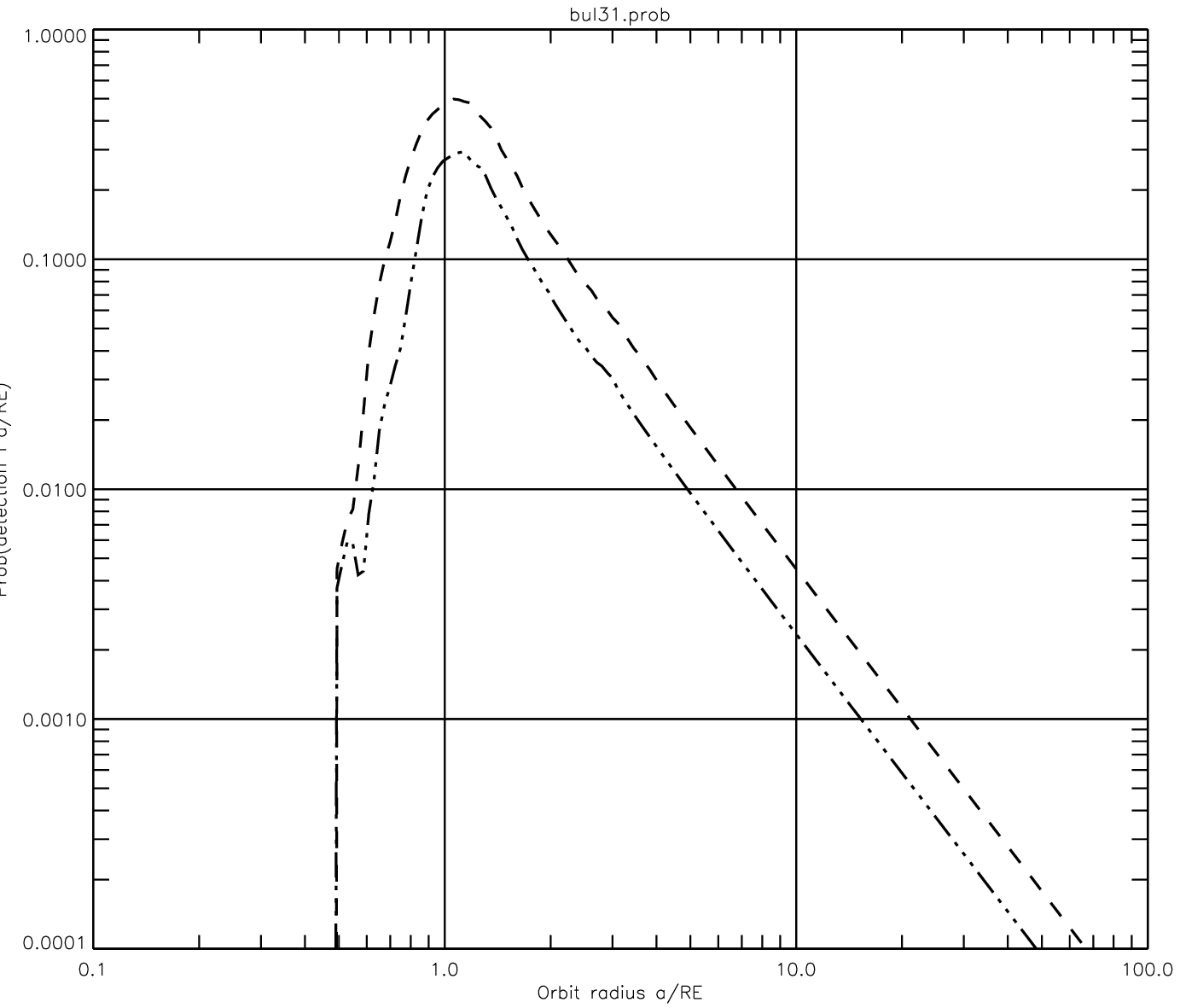,angle=0.0,width=5cm}} \\
\subfigure[2000BUL33]{\label{fig:bul33p}
\psfig{file=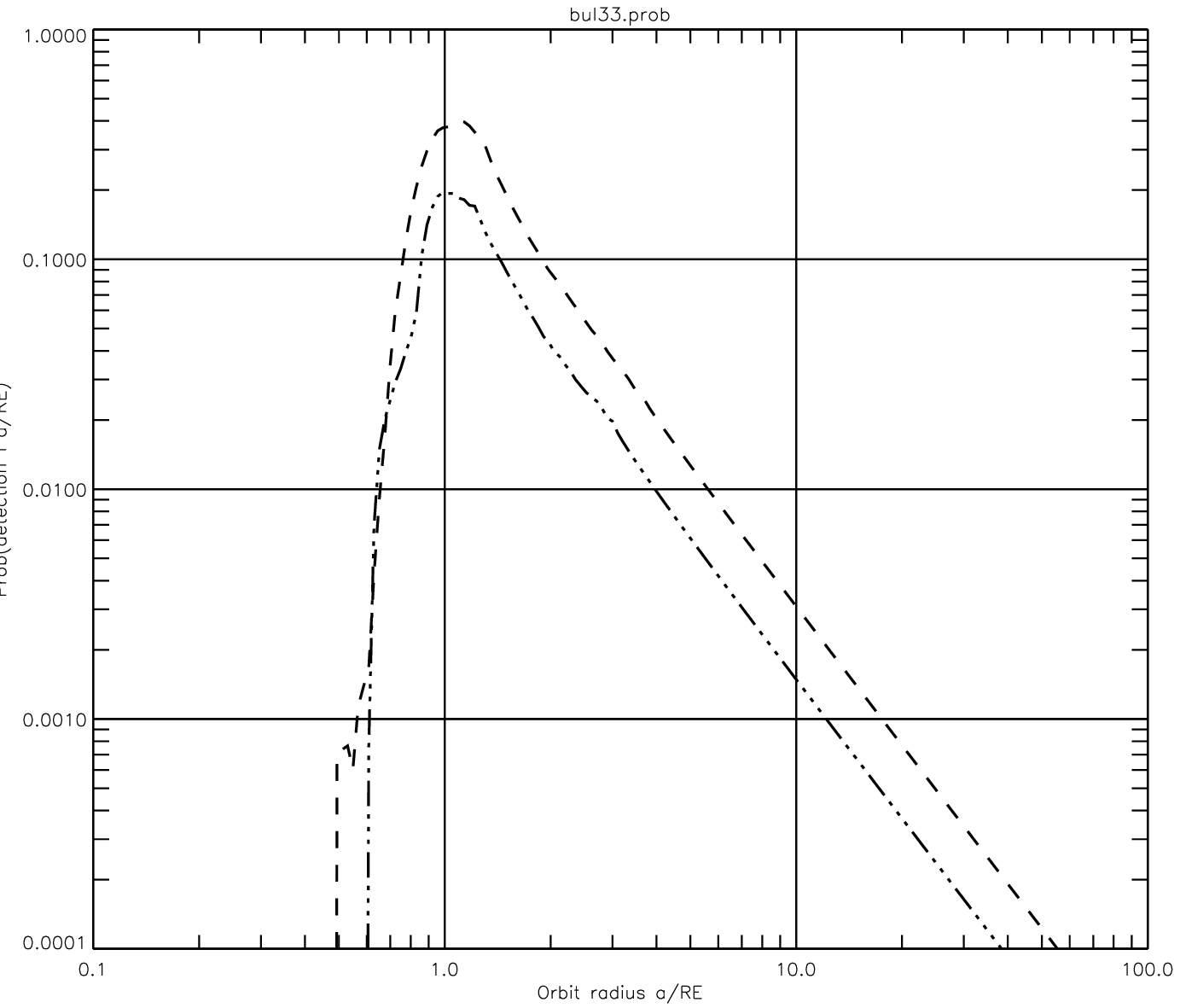,angle=0.0,width=5cm}}
&
\subfigure[2000BUL34]{\label{fig:bul34p}
\psfig{file=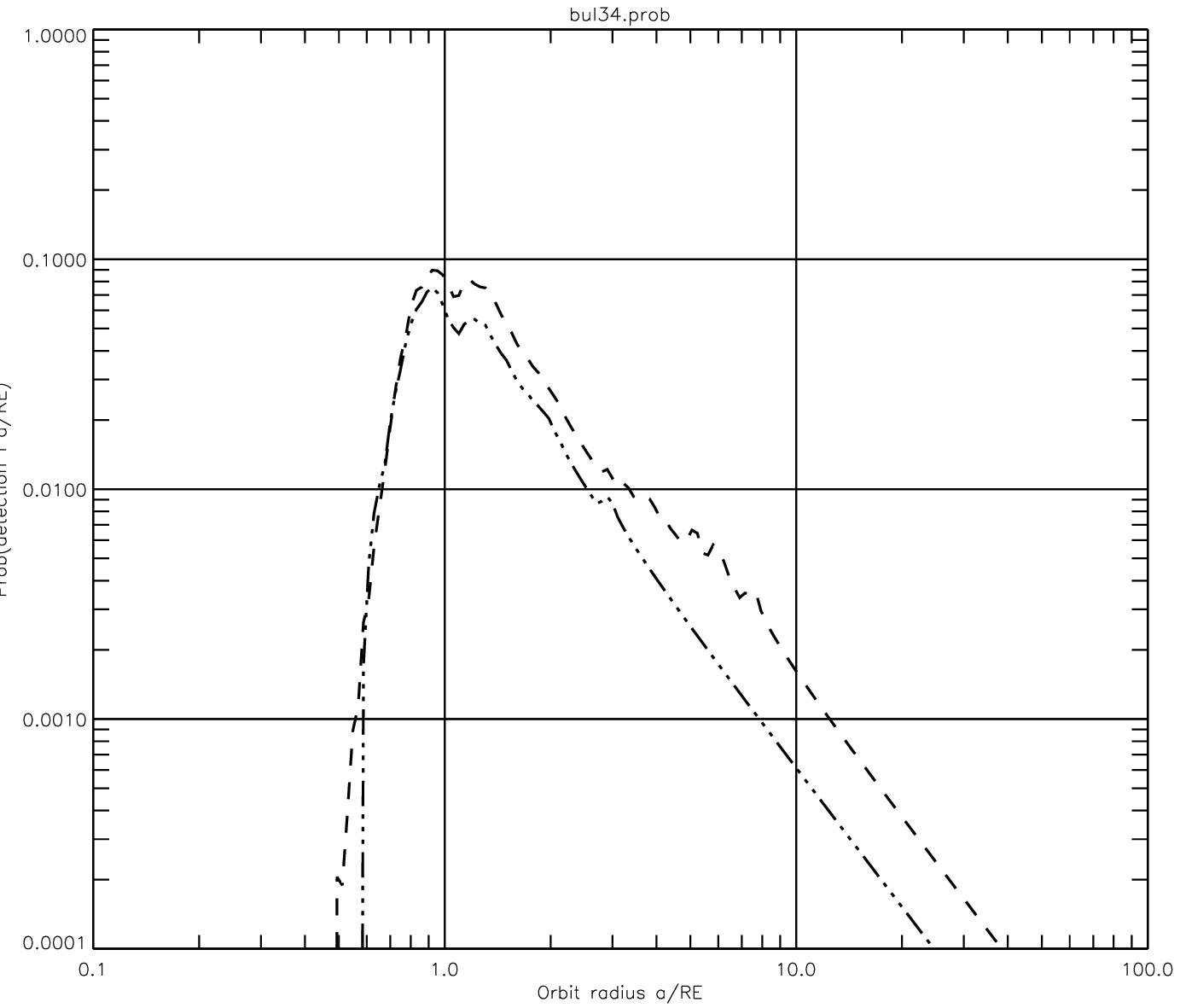,angle=0.0,width=5cm}} 
&
\subfigure[2000BUL36]{\label{fig:bul36p}
\psfig{file=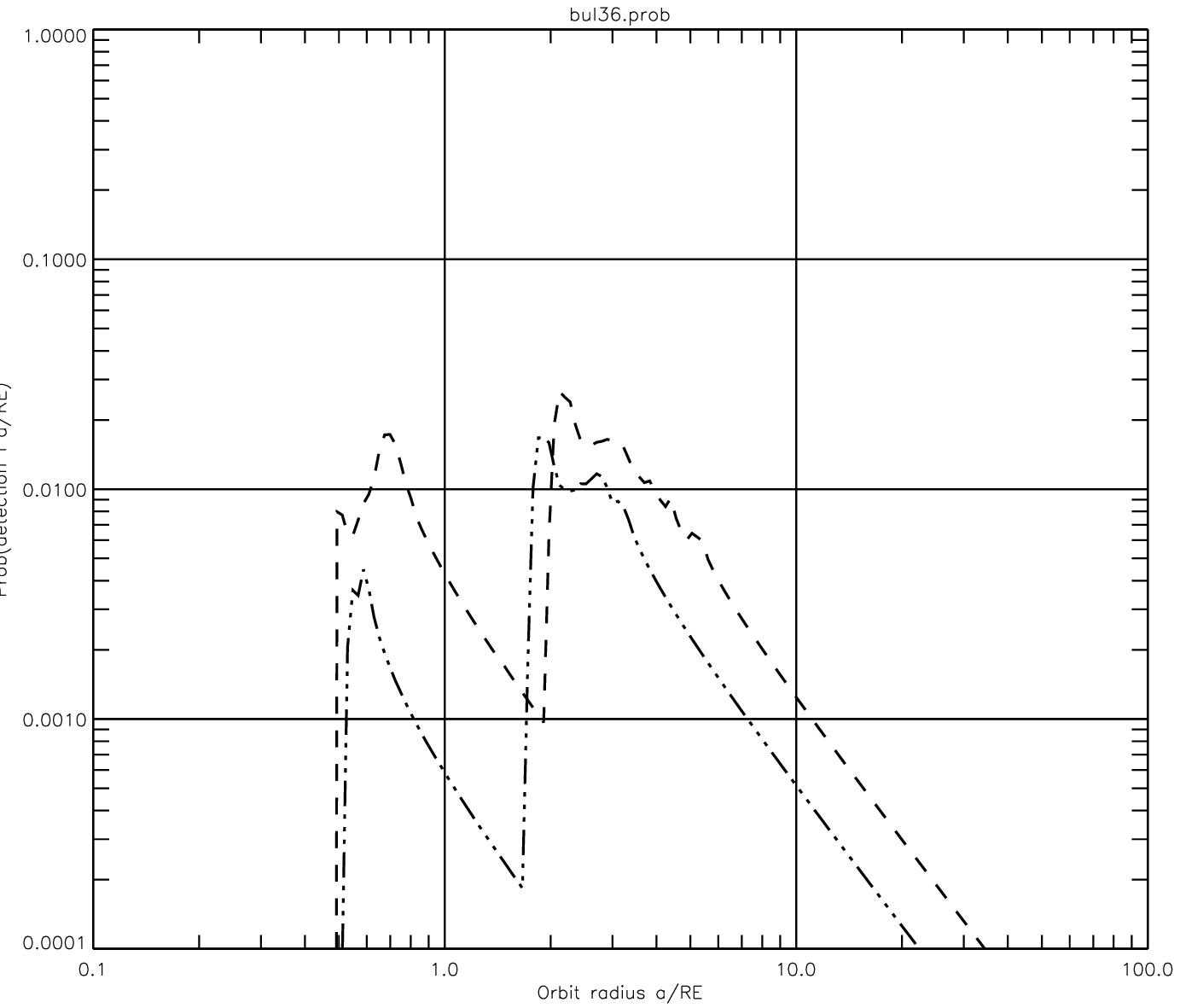,angle=0.0,width=5cm}} \\
\subfigure[2000BUL37]{\label{fig:bul37p}
\psfig{file=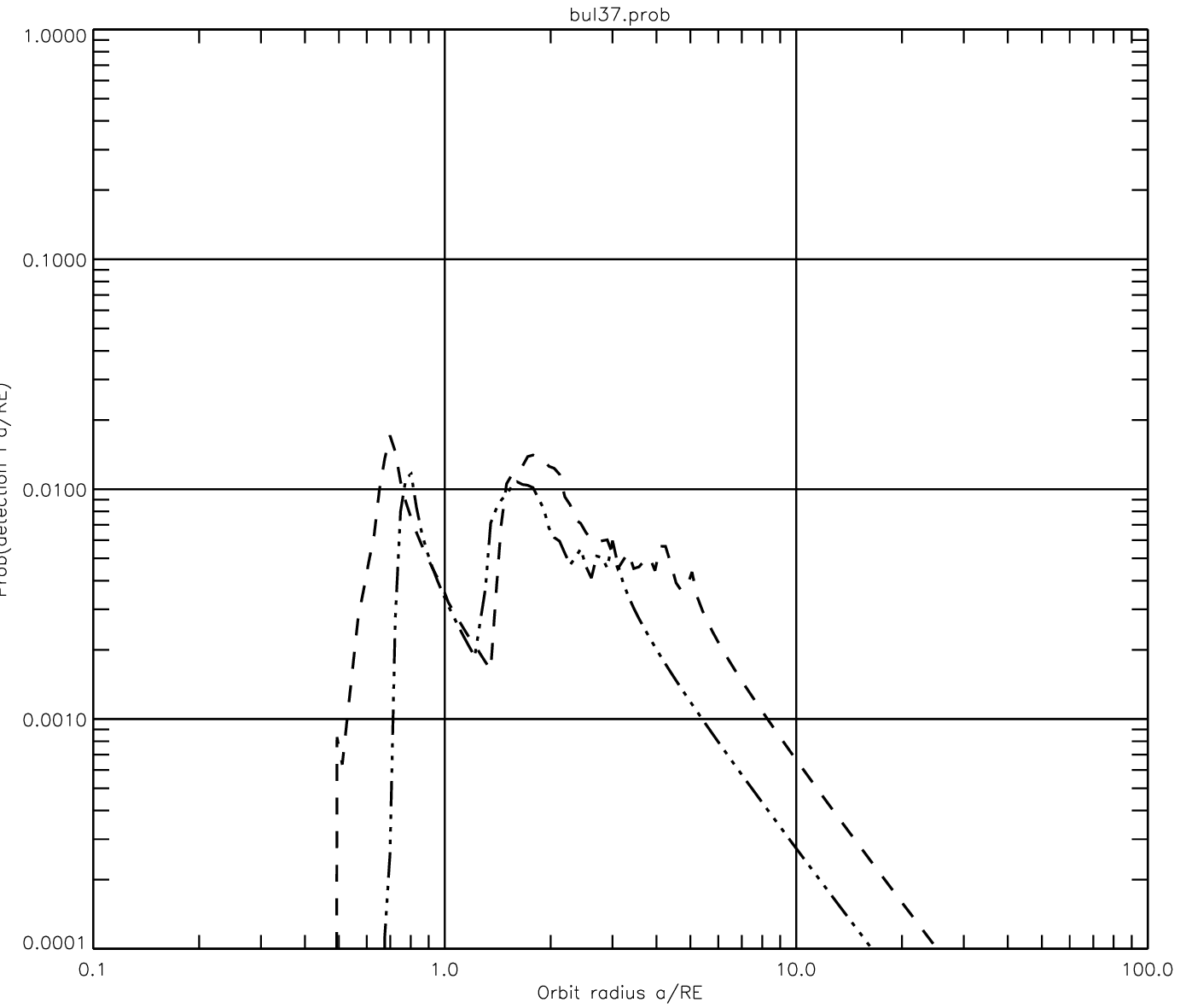,angle=0.0,width=5cm}}
&
\subfigure[2000BUL39]{\label{fig:bul38p}
\psfig{file=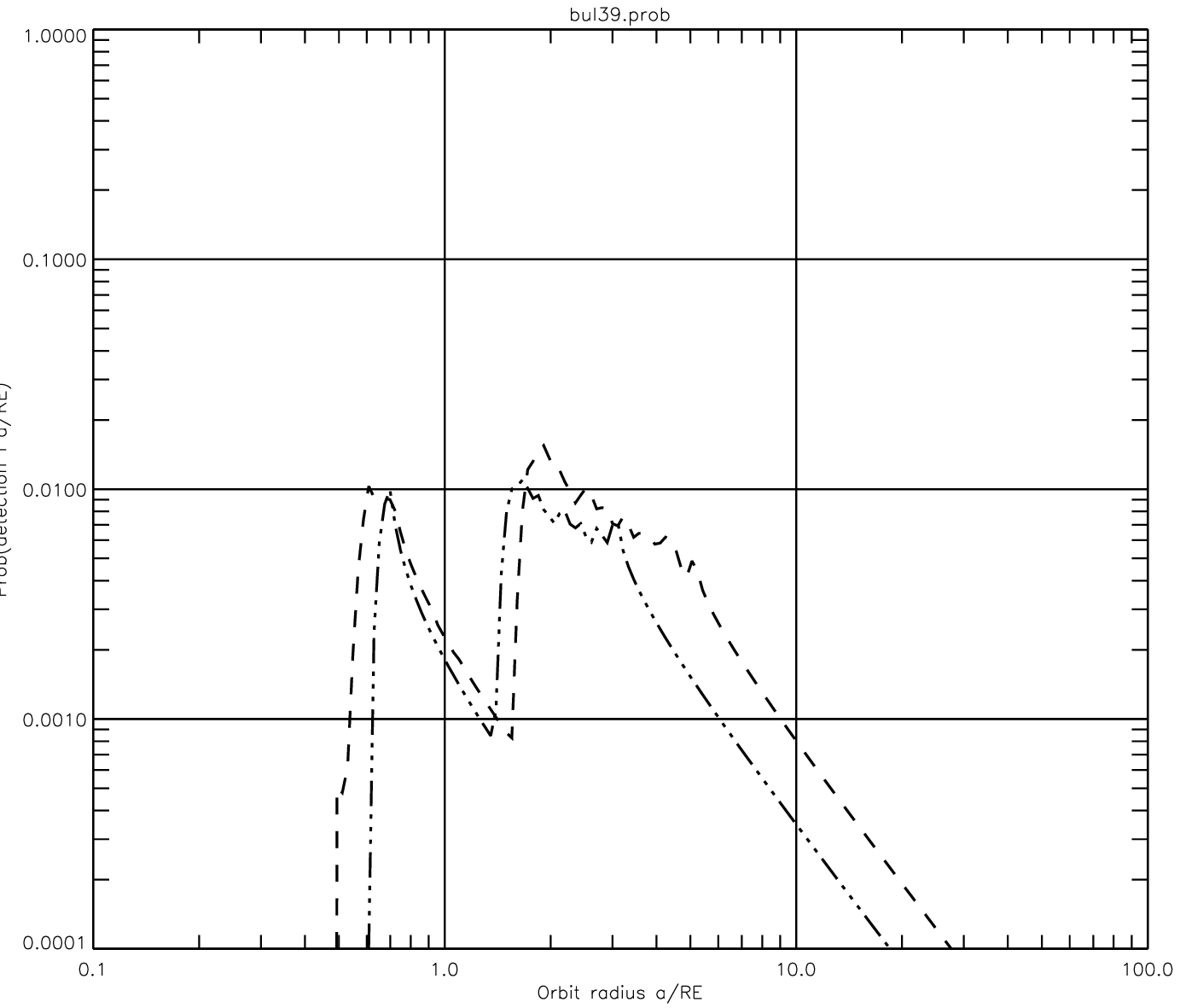,angle=0.0,width=5cm}} 
&
\subfigure[all 8 events]{\label{fig:bulallp}
\psfig{file=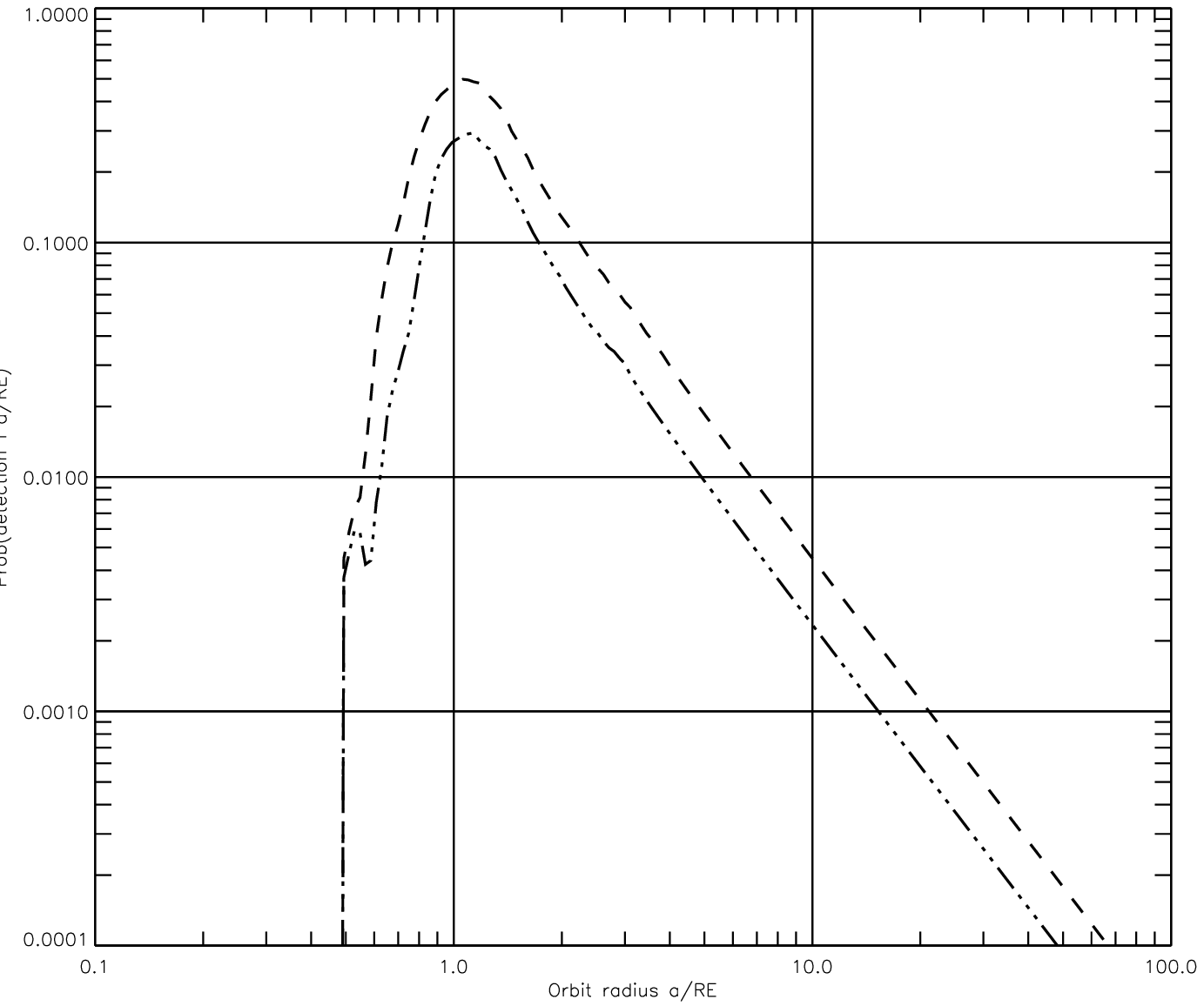,angle=0.0,width=5cm}} \\
\end{tabular}
\caption{Planet detection probabilities for the events observed. The curves are for a planet/star mass ratio $q=10^{-3}$. The results shown are for $\Delta\chi^2$ threshold values of 25 (dashed) and 60 (dot-dashed).}
\protect\label{fig:cpp}
\end{figure*}
\begin{figure*}
\def\subfigtopskip{0pt}
\def\subfigbottomskip{0pt}
\def\subfigcapskip{0pt}
\centering
\begin{tabular}{cc}
\subfigure[2000BUL31]{\label{fig:bul31chi}
\psfig{file=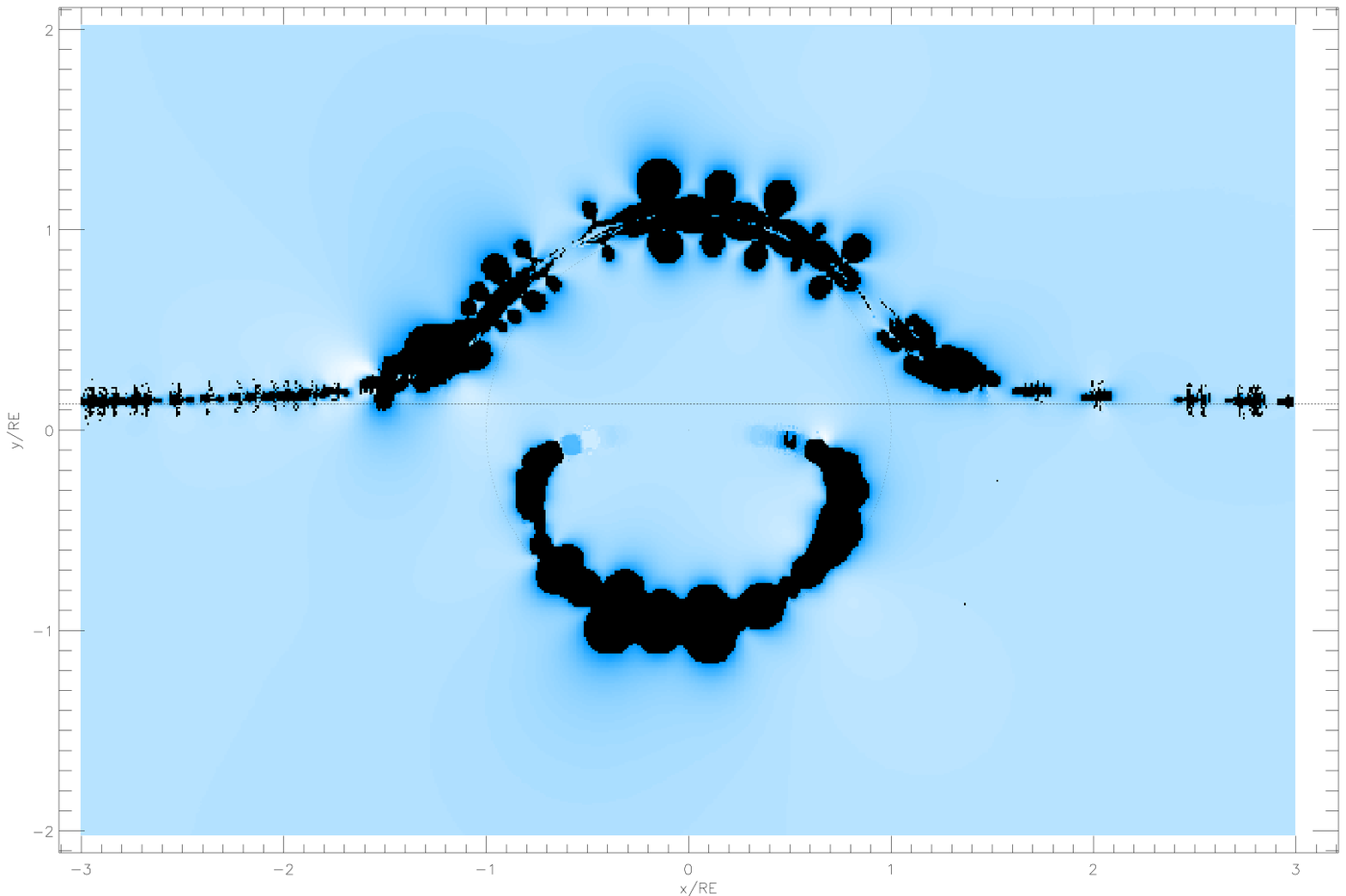,angle=0.0,width=8cm}}
&
\subfigure[2000BUL33]{\label{fig:bul33chi}
\psfig{file=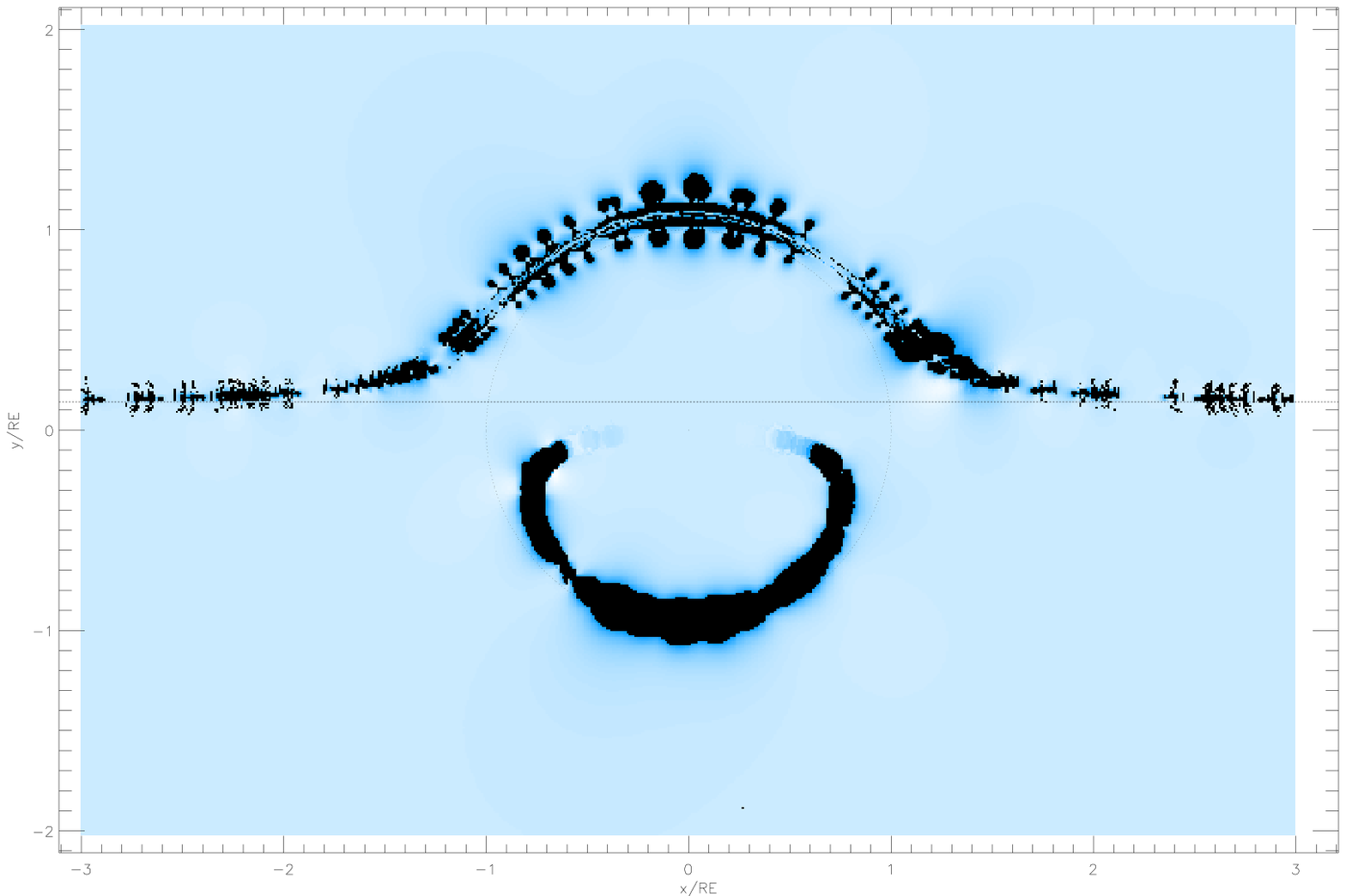,angle=0.0,width=8cm}}
\end{tabular}
\caption{$\Delta\chi^{2}$ -vs- planet position for the data on events 2000BUL31 and 2000BUL33. The black detection zones show where the presence of a planet with a planet/star mass ratio $q = 10^{-3}$ is ruled out by our observations. A successful detection would have been indicated with a white zone. The $\Delta\chi^{2}$ threshold value used for these plots was set at 60.}
\protect\label{fig:chi}
\end{figure*}

\section{POINT-LENS FITS TO LIGHTCURVES}
An effect that we expect to be affecting all lightcurves to some extent is blending. Blending is common in the photometry of crowded fields such as the Galactic Bulge. Its influence might lead to misinterpretation of the baseline magnitude of the source and thus inaccurate estimation of the true maximum amplification and timescale of the lensing event. Blended events may be chromatic so multiband photometry can help estimate this effect \cite{Wozniak97,Vermaak00}.

We performed a combined $\chi^2$ minimization fit to both our dataset and the OGLE publicly available data by optimizing 6 parameters: the time of maximum amplification $t_{0}$, event timescale $t_{\mbox{E}}$, maximum amplification $A_{0}$, baseline magnitude for the OGLE data $I_{\mbox{OGLE}}$,  the magnitude offset $\Delta I = I_{\mbox{OGLE}}-I_{\mbox{JKT}}$ between the two datasets and the blend fraction $b_0 = f_b/f_s$. $f_b$ is the blend flux of unresolved light sources and $f_s$ the flux of the unlensed source. Then the total observed flux at time $t$ becomes $f_{\mbox{tot}}(t) = f_s \times A(t) + f_b$, where $A(t) = (u^2(t) + 2) / ( u(t) \sqrt{u^2(t) + 4} )$. The observed magnification in this case is 
\begin{equation}
A_{obs}(t) = \frac{f_s \times A(t) + f_b}{(f_s + f_b)} = \frac{A(t) + b_0}{(1 + b_0)}.
\end{equation}

Our 6 parameter fits to the lightcurves of the events give $\chi^{2}/(N-6)$ values that are generally higher than their expected value 1 $\pm (2/(N-6))^{1/2}$. To remedy this, we refit the combined dataset by introducing two additional parameters that adjust the error bars, $f$ and $\sigma_{0}$. $f$ respresents a scale factor for the error bars accounting for the under- or over-estimation of the initial error bars. The $\sigma_{0}$ term is an additive flux error intended to account for crowded field effects and is most important when the source is unmagnified.

So the error bar $s_{i}$ on the flux assumes the form:
\begin{equation}
s_{i} = \left( {\sigma_{0}}^2 + f^2 {\sigma_{i}}^2\right)^{1/2}
\end{equation}
where the initial estimates for $\sigma_{0}$ and $f$ are 0 and 1 respectively.
By having added these two extra parameters to adjust the size of the error bars, we cannot use $\chi^2$ minimization to optimize the fit since the fit can achieve $\chi^2 = 0 $ by making the error bars infinitely large. Therefore we use a maximum likelihood criterion. Assuming Gaussian error distributions this is equivalent to minimizing
\begin{equation}
\chi^2 + 2 \displaystyle\sum_{i=1}^{N} {\mbox{ln}}(s_{i}).
\end{equation}
Thereby we introduce a penalty in making the error bars large since that increases the value of the second term in the equation.

Our $\chi^2$ fits to the lightcurves of the events using 8 parameters are summarised in Table \ref{tab:observations} and plots of the fits are presented in Figures \ref{fig:magplots1} and \ref{fig:magplots2}. The residuals of the fits are shown in \ref{fig:resplots1} and \ref{fig:resplots2}. OGLE data are indicated by diamonds and JKT data by squares. Table \ref{tab:observations} gives for each event the number of JKT and OGLE data points fitted and the 8 parameters arising from the fit.

For 2000BUL26 we obtained data during the second half of the lightcurve. The scatter in the data is generally consistent with noise. No significant deviations ($>5\sigma$) from the point-source point-lens (PSPL) lightcurve are seen with the exception of one point at $\sim 14$mag which deviates by $+7\sigma$. We cannot confirm the cause of this deviation since we have no access to the OGLE frames but it is unlikely to be due to a planetary companion since further observations obtained by the OGLE team close to that time do not deviate from the expected values.

Because at zenith distance (ZD) $>$45 degrees in the west the JKT optics show irregular jumps that usually appear as double images on the CCD, half of the observations for event 2000BUL29 had to be rejected as inappropriate for photometric analysis. As a consequence, this event was not well sampled, and we only obtained 5 data points. The fit is therefore mainly constrained by the OGLE data points.

The best sampled events were 2000BUL31, whose decline we missed because of the expiry of our allocated observing time, 2000BUL33 and 2000BUL34 where our data cover a significant portion of the amplification. The good coverage and high amplifications of these events allow us to place strict constraints on the fitted parameters and to exclude the presence of planetary conpanions in the lensing zone ${0.6 \le a/R_{E} \le 1.6}$ ({\it a} being the planetary orbital radius) with high levels of confidence as discussed in section 6.

2000BUL37 was again covered in the decline and we obtained good coverage of the second half of the peak. The OGLE dataset lacks any points in the decline and it is the JKT data that help to define the shape of the lightcurve. 2000BUL36 and 2000BUL39 were low amplification events selected by the priority algorithm mainly because they were close to maximum amplification while the remaining ongoing events at the time were away from their maximum amplification values. Clearly, the information extracted from these last two events is not of the highest quality as their faintness and low amplifications result in poorer data points and a deviation should be more pronounced to be detected. 
All data is available upon request.

\section{EVENT DETECTION PROBABILITIES}
Following the 8-parameter PSPL fit, we refit the data assuming a binary lens \cite{witt90,Schneider86} and proceed to calculate the net detection probability (for a given mass ratio $q$) for each of the sampled events. This involves two additional parameters, $d$ and $q$, where $d$ is the projected separation between the planet and the star, and $q$ the planet to star mass ratio. A similar analysis using a different method has been recently presented in \cite{gaudi00,Albrow01}. 

Prior to calculating the detection probability for a planet of mass ratio $q$ we set up a fine grid of planet positions in $x,y$ on the lens plane and for each of these positions we fit the binary model to the data optimizing all parameters. The density of sampling in $x,y$ has to be dense enough so that planetary fits are not missed. Our grid step size spacing was defined as $\sqrt{q}/4$, where $q$ is the mass ratio. This sets up a very fine grid for each selected mass ratio. We then make a $\Delta\chi^2$ map versus planet position by subtracting the minimum $\chi^2$ of the PSPL fit from the minimum $\chi^2$ of the binary fit for each $x,y$. Examples of such maps are shown in figure \ref{fig:chi}. All $x,y$ values where the $\Delta\chi^2$ exceeds the threshold value (${\Delta\chi^2}_{\mbox{thr}}$=60) are shown in black. These `black zones' show us where the PSPL model gives a better fit to the data. The appearance of `white zones' on the plots would signify a better fit has been achieved with the assumption of a planet at that position interacting with one of the two images of the source star ($\Delta\chi^2 \leq$ $-{\Delta\chi^2}_{\mbox{thr}}$). Our plots show no white detection zones for any of the 8 events followed. Note that the detection zones closer to the Einstein ring of the lens are larger. This is because a planet in that vicinity would be perturbing a more highly amplified image.

We calculated the detection probability for a planet/star mass ratio $q=10^{-3}$ for all 8 events followed using two detection threshold values ${\Delta\chi^2}_{\mbox{thr}}$=25,60. The results are plotted in figure \ref{fig:cpp}. Briefly, the detection probability of finding a planet at the position $x,y$ on the lens plane for an orbital radius $a$ is calculated as:
\begin{equation}
P(det|a) = \int P(det|x,y) P(x,y|a) dx dy.
\end{equation}
The first term on the right side is 
\begin{equation}
P(det|x,y) = 
\begin{cases}
  1& \text{if  $\Delta\chi^{2} > {\Delta\chi^2}_{\mbox{thr}}$ } \\
  0& \text{otherwise}.  \\
\end{cases}
\end{equation}
where $\Delta\chi^{2}$ is the change in $\chi^{2}$ for a planet at $x,y$ relative to the no-planet model. This term becomes significant when the planet at $x,y$ lies close to one of the images of the source at the time of one of the data points in the lightcurve. The second term $P(x,y|a)$ is obtained by randomly orienting the planet's assumed circular orbit of radius $a$, and then projecting it onto the $x,y$ plane of the sky. This gives a circular distribution centred on the lens star and rising as $(d/a)^2$ to a sharp peak at $d=a$, outside which the probability vanishes. This term may be written as:
\begin{equation}
P(x,y|a) = \frac{1}{2 \pi a \sqrt{a^{2}-x^{2}-y^{2}}}
\end{equation}
for ${d^{2} = x^{2} + y^{2} < a^{2}}$.
A slightly elliptical orbit would blur out the outer edge, and it's obviously possible to calculate this for any assumption about the eccentricity.

The net detection probability $P(det|a)$ is therefore the result of summing up the fraction of the time that a planet in the orbit of radius $a$ would be located inside one of the `black zones'.

For our threshold value of ${\Delta\chi^2}_{\mbox{thr}}$=60, the probability of detection reaches a highest value of $\sim 30$\% for $q=10^{-3}$ and $a/R_{E} \sim 1$ for event 2000BUL31 and $\sim 20$\% for 2000BUL33. The remaining events that were followed have lower (1-10\%) detection probabilities. The probability then drops off for larger separations. As a general trend the probability peaks around 1 $R_{E}$ as expected while for events with low amplifications, for the mass ratios presented, we observe a double peaked probability in agreement with the simulations performed by Gaudi \cite{gaudi00} for the detection efficiency. The total probability from all 8 events combined is obtained by summing up the individual probabilities for each event and for each value of $a/R_E$. It is dominated by the contributions from 2000BUL31 and 2000BUL33 and has a maximum value of 63\% for $a \sim R_{E}$ (see Fig. \ref{fig:bulallp}) for ${\Delta\chi^2}_{\mbox{thr}}$=60. This maximum approaches 100\% if we use a detection threshold to 25.

\section{SUMMARY}
We have followed 8 microlensing events using the JKT on La Palma for 2hrs per night from 6 June to 17 July 2000. We presented fits to the combined JKT-OGLE datasets and recalculated the event parameters. We searched the data for signatures of planetary companions with mass ratio $q=10^{-3}$ but have seen no indications of a planetary presence in the datasets. Finally we calculated the planetary detection probabilities on all the events for a mass ratio of $q=10^{-3}$. For events 2000BUL31 and 2000BUL33, our detection probabilities peak at $\sim$30\% and $\sim$20\% respectively for $q=10^{-3}$ and $a \sim R_{\mbox{E}}$ for ${\Delta\chi^2}_{\mbox{thr}}$=60.

\section{ACKNOWLEDGEMENTS}
We would like to thank A. Udalski and the OGLE team for placing their data on these events in the public domain and for providing larger finder charts after our request. Also thanks to Stephen Kane for providing helpful comments on drafts of the manuscript.

\bibliographystyle{mn}
\bibliography{iau_journals,master,Xplanets}

\bsp
\end{document}